\documentclass[12pt,preprint]{aastex}
\usepackage{graphics}
\usepackage{subfigure}
\usepackage{amsmath}
\usepackage{bm}

\shorttitle{Radiation spectra from Langmuir turbulence}
\shortauthors{Teraki et al.}

\begin{document}

\title{GENERAL PROPERTIES OF THE RADIATION SPECTRA FROM RELATIVISTIC ELECTRONS MOVING IN A LANGMUIR TURBULENCE}

\author{Yuto Teraki\altaffilmark{1} and Fumio Takahara\altaffilmark{1}}
\affil{Department of Earth and Space Science, Graduate School of Science, Osaka University, 1-1 Machikaneyama-cho, Toyonaka, Osaka 560-0043, Japan}
\email{teraki@vega.ess.sci.osaka-u.ac.jp}

\begin{abstract}
We examine the radiation spectra from relativistic electrons moving in a Langmuir turbulence 
expected to exist in high energy astrophysical objects by using numerical method.
The spectral shape is characterized by the spatial scale $\lambda$, field strength $\sigma$, 
and frequency of the Langmuir waves, and in term of frequency  
they are represented by $\omega_0 = 2\pi c/\lambda$, $\omega_\mathrm{st}=e\sigma/mc$, and $\omega_\mathrm{p}$, respectively.
We normalize $\omega_\mathrm{st}$ and $\omega_{p}$ by $\omega_0$ as $a \equiv \omega_\mathrm{st}/\omega_0$ 
and $b \equiv \omega_\mathrm{p}/\omega_0$, and examine the spectral shape in the $a-b$ plane.
An earlier study based on Diffusive Radiation in Langmuir turbulence (DRL) theory by Fleishman \& Toptygin showed that 
the typical frequency is $\gamma^2\omega_\mathrm{p}$
and that the low frequency spectrum behaves as $F_\omega \propto \omega^1$ for $b>1$ irrespective of $a$.
Here, we adopt the first principle numerical approach to obtain the radiation spectra in more detail.
We generate Langmuir turbulence by superposing Fourier modes, inject monoenergetic electrons, solve the equation of motion,
and calculate the radiation spectra using Lienard-Wiechert potential.
We find different features from the DRL theory for $a>b>1$.
The peak frequency turns out to be $\gamma^2 \omega_\mathrm{st}$ which is higher than $\gamma^2\omega_\mathrm{p}$
predicted in the DRL theory, and the spectral index of low frequency region is not $1$ but $1/3$.
It is because the typical deflection angle of electrons is larger than the angle of the beaming cone $\sim 1/\gamma$.
We call the radiation for this case "Wiggler Radiation in Langmuir turbulence" (WRL).
\end{abstract}


\section{INTRODUCTION}
\label{int}
The radiation mechanisms of many high energy astrophysical objects are still
an active issue since they often contain features which are not easily explained 
by the conventional synchrotron and inverse Compton emissions.
Recently, much attention has been paid to the radiation signatures from the turbulent electromagnetic fields
(e.g., Medvedev 2000, Fleishman 2006, Kelner, Aharonian, \& Khangulyan 2013, Mao \& Wang 2013, Teraki \& Takahara 2013).
The main scene of the emission regions of high energy astrophysical objects is collisionless shocks, and
the turbulent electromagnetic fields would be generated in the shock region.
Therefore, the electromagnetic turbulence should be taken into account when we consider the radiation.
However, for the major emission mechanisms of synchrotron radiation and inverse Compton scattering,
effects of small scale turbulence are not taken into account.
There is a room for novel emission signatures in the consideration of turbulence, 
which may be of relevance to observations.
By reproducing the observed spectra, we can extract physical parameters of astrophysical objects.
Thus, researches of radiation signatures from the turbulent field would play a key role for the understanding 
of physical mechanisms of the high energy astrophysical objects.

Radiation spectra from a small scale turbulent magnetic field have been well studied 
as the "jitter radiation" or "Diffusive Synchrotron Radiation" (Medvedev et al. 2011, Fleishman \& Urtiev 2010
and references therein).
Differences from the synchrotron radiation become significant when the typical spatial scale of an eddy 
$1/k_\mathrm{typ}$ is smaller than the Photon Formation Length (PFL) of the synchrotron photons $r_\mathrm{L}/\gamma$,
where $r_\mathrm{L}$ is the Larmor radius and $\gamma$ is the Lorentz factor of the electron.
Such a small scale magnetic field is thought to be generated by Weibel instability around the shock front.
When the strength of this small scale turbulent magnetic field is dominant, 
the radiation spectra are determined by the turbulence and reveal various signatures of the turbulent field.
For example, when $2\pi/k_\mathrm{typ}\ll r_\mathrm{L}/\gamma$ 
the peak frequency in $\nu F_\nu$ spectrum becomes $\gamma^2k_\mathrm{typ}c$, and 
the spectrum in the highest frequency region shows a power law $\nu F_\nu\propto \nu^{-\mu+1}$
when the turbulence exists up to the maximum wave number $k_{\rm max}\gg k_{\rm typ}$.
The power law index $\mu$ reflects that of the turbulent magnetic field $B^2(k)\propto k^{-\mu}$.
The spectra show more complex signatures when $2\pi/k_\mathrm{typ}\sim r_\mathrm{L}/\gamma$
(Medvedev 2011, Reville \& Kirk 2010, Teraki \& Takahara 2011).
We note that the anisotropy of turbulence also affects the radiation spectra 
(e.g. Kelner et al. 2013, Reynolds \& Medvedev 2012, Medvedev 2006).

The radiation from an electron which moves in non-uniform magnetic field in a laboratory is well studied using
the insertion device of synchrotron orbital radiation factory,
where a series of magnets are line-upped to make the particle deflect periodically.
It is called "Wiggler" or "Undulator" (Jackson 1999).
For Undulator, the strength of magnets $B$ and gaps between them $\lambda$ are
chosen to satisfy the condition that the observer is always in the beaming cone.
On the other hand, for Wiggler, the observer is periodically in and off the beaming cone.
We estimate the critical distance $\lambda_{\rm c}$ which divides Wiggler and Undulator.
The deflection angle in one deflection is $\theta_{\rm def}= \lambda/r$,
where $r\simeq \gamma mc^2/eB$ is the typical curvature radius of the orbit. 
The radiation from a relativistic particle is concentrated into small cone with opening angle $\sim 1/\gamma$.
Therefore, the critical condition dividing Wiggler and Undulator is $\theta_{\rm def} = 1/\gamma$,
which is rewritten as $\lambda_{\rm c} = r/\gamma$.
Thus, the device is called Undulator when $\lambda<\lambda_{\rm c}$, while
it is called Wiggler when $\lambda>\lambda_{\rm c}$.
The radiation spectrum of Undulator shows a sharp peak at $\gamma^2 2\pi c/\lambda$, while Wiggler 
shows a broad spectrum with peak frequency $\sim \gamma^2 eB/mc$.
The relation between typical frequencies and deflection angle is a key point for 
understanding of the radiation spectra.
Perturbative jitter radiation or perturbative DSR is recognized as extensions of the Undulator radiation,
since the spatial scale of turbulence $\lambda$ is assumed to be much smaller than $mc^2/eB$.

The fact that $\theta_{\rm def}$ determines the radiation feature
was also noted in astrophysics in early seventies by, e.g., Rees, (1971), and Getmantsev \& Tokarev (1972).
Rees assumed that the strong electromagnetic wave is emitted from the Crab pulsar with frequency $\Omega$ (30Hz)
according to the "oblique rotator" model by Ostriker \& Gunn (1969),
and argued that the radiation from an electron moving in the strong wave is not 
inverse Compton scattering (with frequency $\gamma^2\Omega$) but synchrotron-like (synchro-Compton) radiation,
because the deflection angle $\theta_{\rm def}$ is estimated as $\theta_{\rm def}>10/\gamma$,
where the factor $10$ comes from the ratio of cyclotron frequency to the wave frequency $eB_{\rm eq}/mc\Omega$.
The magnetic field strength $B_{\rm eq}$ is estimated by equating the spindown luminosity and power of magnetic dipole radiation
of the Crab pulsar and by considering the distance from the pulsar.
He argued that the radiation signature of synchro-Compton radiation resembles that of the synchrotron radiation.
The typical frequency is determined by the field strength as $\gamma^2eB_{\rm eq}/mc$,
and the low frequency spectral index is roughly $1/3$.
Getmantsev \& Tokarev (1972) studied the radiation spectra under general electromagnetic fields and
stated that the radiation signature from a single charged particle is determined 
by the frequency or wavelength of the field for $\theta_{\rm def}\ll1/\gamma$,
while it is determined by the field strength for $\theta_{\rm def}\gg1/\gamma$ in the same way as synchrotron radiation.
Note that they did not discussed explicit expressions of the radiation spectra from a single electron
for $\theta_{\rm def}\sim 1/\gamma$, they presented a general expression of radiation spectrum
from an ensemble of electrons with a power spectrum.

Thirty years after these pioneering works,
radiation spectra from a relativistic particle interacting with turbulent fields are gathering a renewed attention.
For example, the radiation mechanism from Langmuir turbulence treated in the present paper has become an interesting topic.
The Langmuir turbulence has been pointed out to be generated around the shock front of the relativistic shocks 
(Silva 2006, Dieckmann 2005, Bret, Dieckmann, \& Deutch 2006),
so it should be as important as the radiation from turbulent magnetic field.
Fleishman \& Toptygin (2007a,b) have made a systematic treatment of diffusive radiation in Langmuir turbulence
(see references cited in Fleishman \& Toptyigin 2007b for other earlier relevant works).
Their method is the most sophisticated treatment for Langmuir turbulence, which is based on Toptygin \& Fleishman (1987).

For later discussions, we shortly review their treatment.
They treat the electron motion by a statistical approach and use a perturbative treatment for calculation of the radiation.
The calculation formula for the radiation spectra is based on the one written in Landau \& Lifshitz (1971).
For $\theta_{\rm def}\ll 1/\gamma$, a rectilinear trajectory with constant velocity is assumed,
but non-zero acceleration from the external field is taken into account.
The wavenumber of the Langmuir waves is assumed to satisfy the condition $k_{\rm typ} <\omega_{\rm p}/c$,
where $\omega_\mathrm{p}$ is the plasma frequency.
They calculate the correlation between the acceleration and the Langmuir waves.
The peak frequency is $\gamma^2\omega_\mathrm{p}$.
The spectrum shows an abrupt cutoff above the peak, and becomes $F_\omega \propto \omega^{-\mu}$
in higher frequencies when the turbulence exists up to the maximum wavenumber $k_{\rm max}\gg c/\omega_{\rm p}$
for a power law turbulent spectrum  $E^2(k)\propto k^{-\mu}$.
The spectrum just below the peak is $F_\omega\propto \omega^1$,
and becomes $F_\omega \propto \omega^0$ in lower frequencies and $F_\omega \propto \omega^{1/2}$
in even lower frequency region.\footnote{$F_\omega \propto \omega^2$ spectrum is predicted
in the lowest frequency region, which comes from the effect of wave dispersion in the plasma.
We do not discuss such effects in this paper.}
$F_\omega \propto \omega^{1/2}$ spectrum comes from the effect of multiple scattering.
The angle between the velocity and observer direction becomes larger than the beaming cone after many deflections
even when the deflection angle in one deflection $\theta_{\rm def}$ is much smaller than $1/\gamma$, 
and the approximation of rectilinear trajectory is broken.
Therefore, this treatment is beyond the perturbative treatment, and they call it "non-perturbative treatment".
They treat the changing of direction of motion in many deflections by diffusion approximation.
In consequence, a spectral break emerges in the low frequency region with a suppression of low frequency photons. 
The spectrum becomes $F_\omega \propto \omega^{1/2}$ from this effect and the index of $1/2$ comes from the diffusivity.
They claimed that the break frequency approaches the peak frequency as $\theta_{\rm def}$ becomes large
and the break and peak merge for $\theta_{\rm def}\sim 1/\gamma$.
They stated that even for $\theta_{\rm def}\gg1/\gamma$the spectrum in frequency region just below the peak of 
$\gamma^2\omega_{\rm p}$ is $F_\omega \propto \omega^{1/2}$.
This is inconsistent with the statement by Getmantsev \& Tokarev (1972).
This is one of the objectives of investigations in the present paper.
The effect of large angle deflection would come into play in forming the radiation spectra 
for Langmuir turbulence as in the Wiggler radiation for $\theta_{\rm def}>1/\gamma$.
This point has been discussed for magnetic turbulence in Kelner et al. (2013), Medevev et al. (2011), and Teraki \& Takahara (2011).
In this paper, we investigate general properties of the radiation spectra from relativistic electrons in a Langmuir turbulence
for various cases including this regime.

Before proceeding to the formulation of the calculation employed in the present paper, 
we point out the differences between the magnetic field generated 
by Weibel instability (entropy modes) and the Langmuir waves.
The first is time variability.
The entropy mode does not oscillate, so the typical timescale is determined by the turnover time of an eddy.
It is longer than the crossing time of a relativistic electron if we assume that the background plasma is sub-relativistic.
Therefore, we can treat the magnetic field as a static field when we calculate the radiation spectra for the zeroth-order approximation.
On the contrary, we should not treat the Langmuir turbulence as a static field even for the zeroth-order approximation.
The crossing time can be comparable to or longer than the period of the Langmuir waves, 
because the typical spatial scale is about inertial length $c/\omega_{\rm p}$,
governed by the plasma frequency $\omega_\mathrm{p}$ (Diekmann 2005).
Therefore, the time variability of the electric field can not be ignored.
This effect is well studied by Fleishman \& Toptygin (2007a,b).
The second is the energy change of the radiating electrons.
For the case of turbulent magnetic field, the energy of electrons is conserved if we ignore the radiation back reaction.
However, the energy change cannot be ignored for the Langmuir turbulence,
because the electric field can accelerate the electrons parallel to their velocity.
The Lorentz factor of the electron can change in a short time by strong Langmuir waves.
This effect may play a role for calculation of the radiation spectra.

The calculation of electron trajectory for the strong turbulence by analytical approach is hard to perform.
Thus, we calculate the radiation spectra by numerical approach from first principle.
We calculate radiation spectra for a wide range of the field parameters.
We study about three factors, the scale length, time variability, and strength.
By sweeping the parameter plane, we generally investigate the radiation spectra 
from a relativistic electron moving in Langmuir turbulence.
In section 2, we describe calculation method, and we show the results in section 3.
In section 4, we give the physical interpretations of the discovered spectral features
using radiation for a spatially uniform plasma oscillation.
In section 5, we make a summary and some discussions.

\section{FORMULATION}
We use Lienard-Wiechert potential directly to calculate the radiation spectra.
It is the same approach as employed in Teraki \& Takahara 2011 for magnetic turbulence.
This method is computationally expensive because the time step is restricted by observed frequency
and the integration time has to be longer than the Photon Formation Time (PFT).
PFT is defined as the coherence time for forming a photon 
with a given frequency (Reville \& Kirk 2010, Akhiezer \& Shul'ga 1987).
Therefore, PFT is a function of the frequency.
Since PFT is $\gamma^2$ times larger than the inverse of the observed frequency
and since the time step should be shorter than PFT, 
the integration time is at least $\gamma^2$ times larger than the inverse of the observed frequency.
As a result, large number of time steps are needed to calculate radiation spectra for highly relativistic case.
Although less heavy approach is proposed by Reville \& Kirk 2010 to overcome this problem by using appropriate approximations,
we use the first principle method we show below, because it does not include any approximations in the calculation formula
of the radiation spectra from prescribed trajectory.
Therefore, this method is suitable for studying the cases for which any approximations are difficult to adopt.
By using moderately relativistic particles, we reduce the computational expenses.
We neglect radiation back reaction of electrons, because the strength of electric field
is sufficiently weaker than $m^2c^4/e^3$ (cf. Jackson 1999).

We use 3D isotropic Langmuir turbulence in this paper.
If the Langmuir waves are generated near the shock front, they may be highly anisotropic on the spot.
However, we assume the isotropic turbulence for two reasons.
First, the Langmuir turbulence in large part of the emission region would be isotropic, since
Langmuir waves interact with background ions and form the Kolmogorov type isotropic turbulence (Treumann \& Baumjohan 1997).
Second, the radiation spectra for 3D and 1D turbulence are not so different except for a particular case that 
the direction of radiating electrons and the wave vector are parallel.
For this case, the radiation is linear acceleration emission.
However, emission from the perpendicular acceleration dominates 
as long as the direction differs from the wave vector direction even slightly (Fleishman \& Toptygin 2007a). 

If the shock is ultra-relativistic, the downstream plasma may be relativistically hot.
For simplicity, we assume a mildly relativistic shock, so that the downstream plasma is sub-relativistic.
We ignore the thermal velocity of background plasma in the dispersion relation of Langmuir waves 
$\omega^2 = \omega_\mathrm{p}^2+3/2k^2 v_\mathrm{e,th}^2$.
Thus, we use the propagating Langmuir waves with the same frequency $\omega_\mathrm{p}$.

We generate 3D isotropic Langmuir turbulence by using Fourier transform description, 
which is slightly modified from the description for magnetic turbulence developed by Giacalone \& Jokipii (1999).
It is described by superpositions of $N$ Fourier modes, each with random phase, and direction
\begin{equation}
  \bm{E}(x) = \sum_{n=1}^N A_n \cos\bigl\{(\bm{k}_n \cdot \bm{x} -\omega_\mathrm{p}t + \beta_n)\bigr\}\frac{\bm{k_n}}{|k_n|}.
\end{equation}
Here, $A_n$, $\beta_n$, $\bm{k_n}$, and $\omega_\mathrm{p}$ are the amplitude, phase, wave vector,
frequency of the $n$th mode, respectively.
The amplitude $A_n$ of each mode is defined as  
\begin{equation}
 A_n^2 = \sigma^2 G_n \Biggl[\sum_{n=1}^N G_n\Biggr]^{-1},
\end{equation}
where the variable $\sigma$ represents the amplitude of turbulent field.
We use the following form for the power spectrum 
\begin{equation}
  G_n = \frac{4\pi k_n^2 \Delta k_n}{1 + (k_n L_\mathrm{c})^\mu},
\end{equation}
where $L_\mathrm{c}$ is the correlation length of the field.
Here, $\Delta k_n$ is chosen such that there is an equal spacing in logarithmic $k$-space,
over the finite interval $k_\mathrm{min}\leq k\leq k_\mathrm{max}$ and $N=10^3$,
where $k_\mathrm{min} = 2\pi/L_\mathrm{c}$, $\mu=9/2$ and
$k_\mathrm{max}$ is chosen to be $10^3k_\mathrm{min}$ or $10k_\mathrm{min}$.
The spectrum has a peak at $k_\mathrm{min}$ and the spectral index is  
for 3-dimensional isotropic Langmuir turbulence.

Then we define two parameters which characterize radiation spectra.
The first one is 
\begin{equation}
  a\equiv \frac{e\sigma}{mc^2k_\mathrm{min}}=\frac{\omega_\mathrm{st}}{\omega_0},
\end{equation}
 where $\omega_\mathrm{st}\equiv e\sigma/mc$ and $\omega_0 \equiv k_\mathrm{min}c$.
We call $\omega_\mathrm{st}$ "strength omega", and $\omega_0$ "spatial omega".
The strength omega $\omega_\mathrm{st}$ accounts for the effect of the field strength.
The meaning of strength omega can be understood by considering the curvature of orbit and beaming effects for the relativistic particles
as follows.
For $\gamma\gg1$, the local curvature radius of the orbit suffering from perpendicular acceleration by the electric field 
is $\sim \gamma m c^2/(e\sigma)$.
The radiation is concentrated in the beaming cone with an angle $\sim 1/\gamma$, so the searchlight sweeps
the observer in the timescale of $mc/e\sigma=1/\omega_\mathrm{st}$.
It is an analogy of the cyclotron frequency in the mechanism of synchrotron radiation.
Therefore, $\omega_{\rm st}$ represents the effect of the field strength on the radiation spectra.
As for the spatial omega $\omega_0$, 
since the electron moves nearly at the light speed, the rate of change 
in the force direction for the electron is $2\pi c/\lambda=\omega_0$.
The ratio $\omega_\mathrm{st}/\omega_0=a$ parametrizes the field feature, which 
represents the ratio of the deflection angle $\theta_{\rm def}$ to $1/\gamma$.
Therefore, the spectra from the turbulent magnetic field changes for $a\lesssim1$ to $a\gtrsim1$ drastically 
(Kelner et al. 2013, Medvedev et al. 2011, and Teraki \& Takahara 2011).
Note that $a$ is well known as the "strength parameter".
It means the change of the Lorentz factor of an electron which is accelerated 
along the electric field on the spatial scale $l=1/k_\mathrm{min}$.
We can understand it clearly by using the work by an electric field on an electron as
\begin{equation}
eEl = \Delta \gamma mc^2,
\end{equation}
where we assume the electric field is parallel to the velocity.
We should note that it is different from the one we used in Teraki \& Takahara 2011 by a factor of $2\pi$,
and it agrees with the definition of Fleishman \& Toptygin (2007a,b).

The second one is the ratio of the frequency of the Langmuir waves to $\omega_0$,
\begin{equation}
  b\equiv\frac{\omega_\mathrm{p}}{k_\mathrm{min}c}=\frac{\omega_\mathrm{p}}{\omega_0}.
\end{equation}
For $b\gg1$, the force direction changes with frequency $\omega_\mathrm{p}$ for all electrons.
For $b\ll1$, a change of the force direction is mainly from the spatial fluctuation.
Summarizing above, although there are three parameters of the turbulent field ($\omega_0$, $\omega_\mathrm{st}$, $\omega_\mathrm{p}$),
we can reduce them to two parameters of ($a,b$) when our interest is in the spectral signature.
We investigate the spectral features in the parameter plane of ($a,b$).

We inject monoenergetic electrons with Lorentz factor $\gamma_\mathrm{init}=10$ into the generated turbulent field.
The number of electrons used for each calculation is more than $10^2$ and is written in the caption of each figure.
The initial velocity is randomly chosen to achieve an isotropic distribution.
Next we solve the equation of motion
\begin{equation}
  \frac{d}{dt}(\gamma m \bm{v}) = -e \bm{E},
\label{eom}
\end{equation}
by using the method which has second order accuracy for each time step.
The electrons get energy in the Langmuir turbulence from the parallel (to the velocity) component.
If we pursue electrons much longer time than the PFT of a given frequency, 
the radiation spectrum in high frequency region corresponds to an "integrated spectrum".
This spectrum would be understood by a superposition of "instantaneous spectra".
Here, the term of "instantaneous spectrum" means a spectrum which does not contain the effect
of secular energy change of the radiating electrons.
If we do not obtain the instantaneous spectra, we can not discern either
the intrinsic feature of instantaneous spectra or energy change of radiating electrons determines the spectral features.
Therefore, in this paper we concentrate on "instantaneous spectra".
On the other hand, spectral broadening due to finite integration time of particle orbit is inevitable.
To compromise the accuracy and the instantaneousness, we choose the integration time as $100$ times PFT 
of the peak or break frequency.
Let us take an example of the spectrum for well known jitter radiation\footnote{We use the term "jitter radiation"
when the orbit is rectilinear and acceleration is treated perturbatively whether acceleration is due to
magnetic turbulence or electric turbulence.}, for $a\ll1, b=0$.
The break frequency of the jitter radiation is $2\gamma^2\omega_0$, and corresponding PFT is $T\sim 1/\omega_0$.
We can understand the origin of this frequency by using "the method of virtual quanta", (cf. Jackson 1999).
The spatial fluctuation of the electric field with a spatial scale $\lambda= 2\pi c/\omega_0$.
This fluctuation can be regarded as a photon which has the frequency with $\gamma \omega_0$ in the electron rest frame.
The electron scatters the photon, and transformed back to the observer frame, 
the frequency of the scattered photon is written as $\sim 2\gamma^2\omega_0$.
In other words, we can understand this process as an analogy of the inverse Compton scattering for the photon with frequency $\omega_0$.
We note that this analogy is based on the assumption that the observer is in almost the center of the beaming cone over the PFT.
It is valid when $a\ll1$.
The deflection angle in the time scale of $1/\omega_0$ is $eE\lambda/\gamma mc^2$,
and when we require that it is smaller than $1/\gamma$, we get the condition $eE\lambda/mc^2<1$.
It can be transformed to $\omega_\mathrm{st}/\omega_0=a<1$.
An alternative explanation can be done by using  "the photon chasing effect" (Rybiki \& Lightmani 1979).
Since the radiation from an ultra-relativistic particle is concentrated into the beaming cone, the particle chases the photon.
Therefore, the observer sees the emitted radiation in a time span $(1-v/c)T$, 
so the frequency is $1/(1-v/c)T\sim 2\gamma^2/T \sim \gamma^2\omega_0$.
As a consequence, the typical frequency is $2\gamma^2\omega_0$.
As we see in the next section, PFT for the peak frequencies of the spectra is 
the one of the typical timescales of $1/\omega_0$, $1/\omega_\mathrm{st}$, and $1/\omega_\mathrm{p}$.
We choose a suitable one for each case.
The integration time $100$ times the PFT of the peak frequency 
is sufficiently long to resolve the spectral shape in the low frequency regions.

We calculate the radiation spectra using the acceleration, velocity, and position of electrons.
The energy $dW$ emitted per unit solid angle $d\Omega$ (around the direction $\bm{n}$) 
and per unit frequency $d\omega$ to the direction $\bm{n}$ is computed as
\begin{equation} 
  \frac{dW}{d\omega d\Omega} = \frac{e^2}{4 \pi c^2} 
  \Bigl| \int^{\infty}_{-\infty} \:dt^{\prime} \frac{ \bm{n} \times \bigl[ (\bm{n} - \bm{\beta}) \times \dot{\bm{\beta}} \bigr] } 
  {(1 - \bm{\beta} \cdot \bm{n} )^2 }\exp\bigl\{{i\omega ( t^{\prime} - \frac{\bm{n} \cdot \bm{r}(t^{\prime})}{c})}\bigr\} \Bigr|^2,
\end{equation}
where $\bm{r}(t^{\prime})$ is the electron trajectory, $t^{\prime}$ is retarded time (Jackson 1999). 
Since we have not any specific constraints for $a$ and $b$ which are realized in the high energy astrophysical object,
we sweep wide parameter range of $a$ and $b$.

\section{NUMERICAL RESULTS}

\subsection{Short wavelength regime}
\begin{figure}[htb]
  \begin{center}
    \subfigure[$a=0.1$]{
      \includegraphics[width=.45\columnwidth]{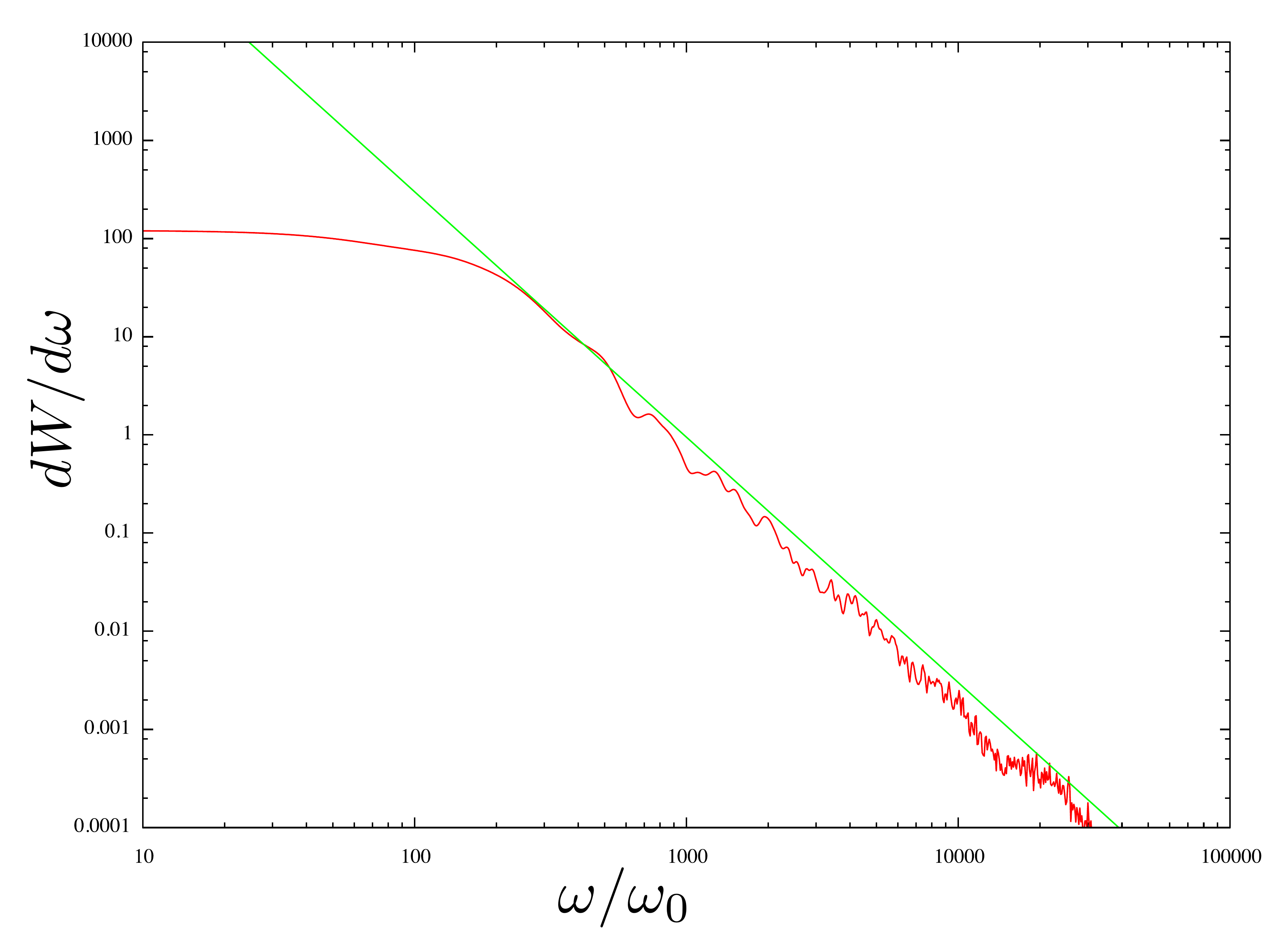}   
    }~
    \subfigure[$a=5$]{
      \includegraphics[width=.45\columnwidth]{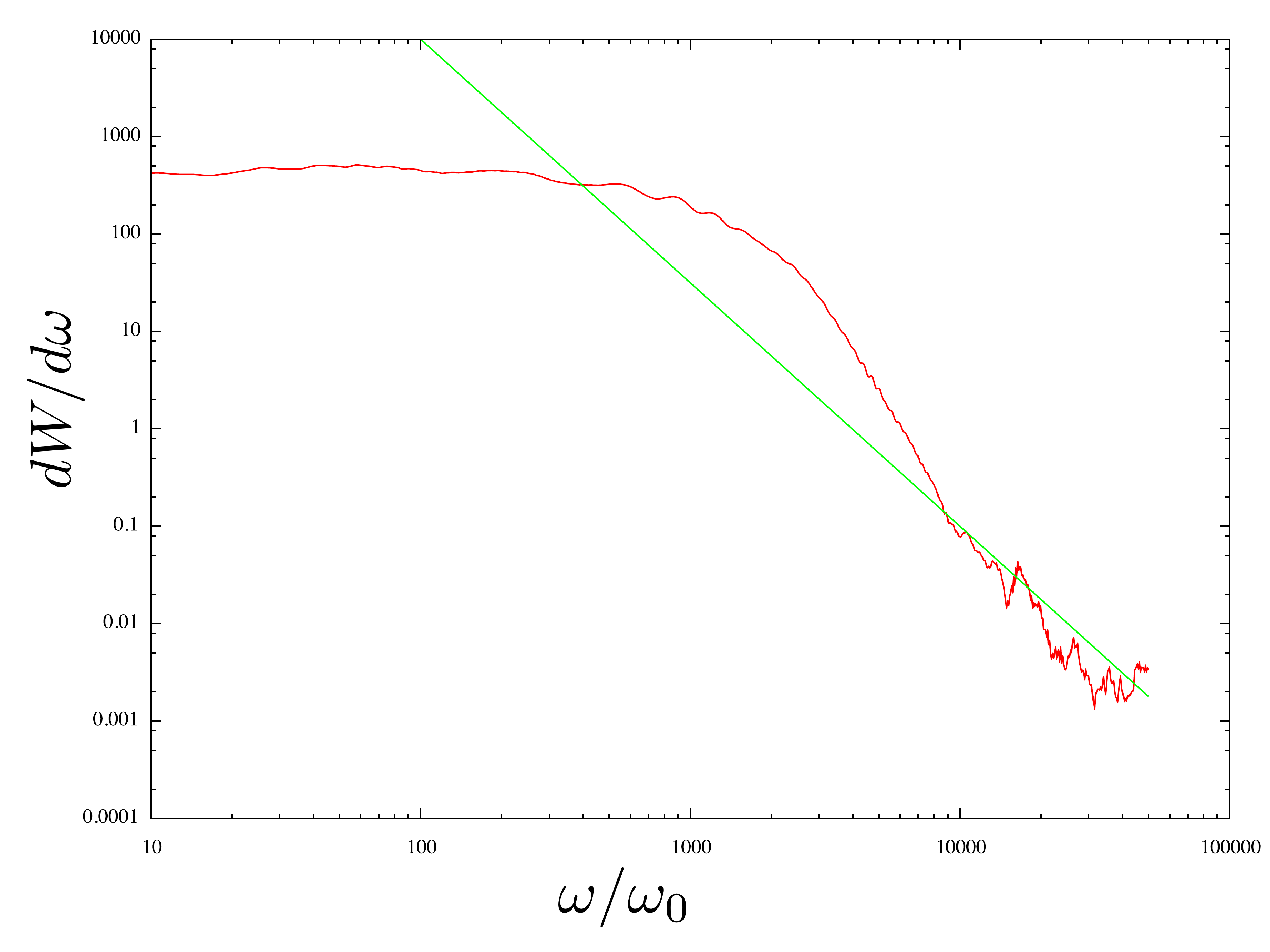}
    }\\
    \subfigure[$a=10$]{
      \includegraphics[width=.45\columnwidth]{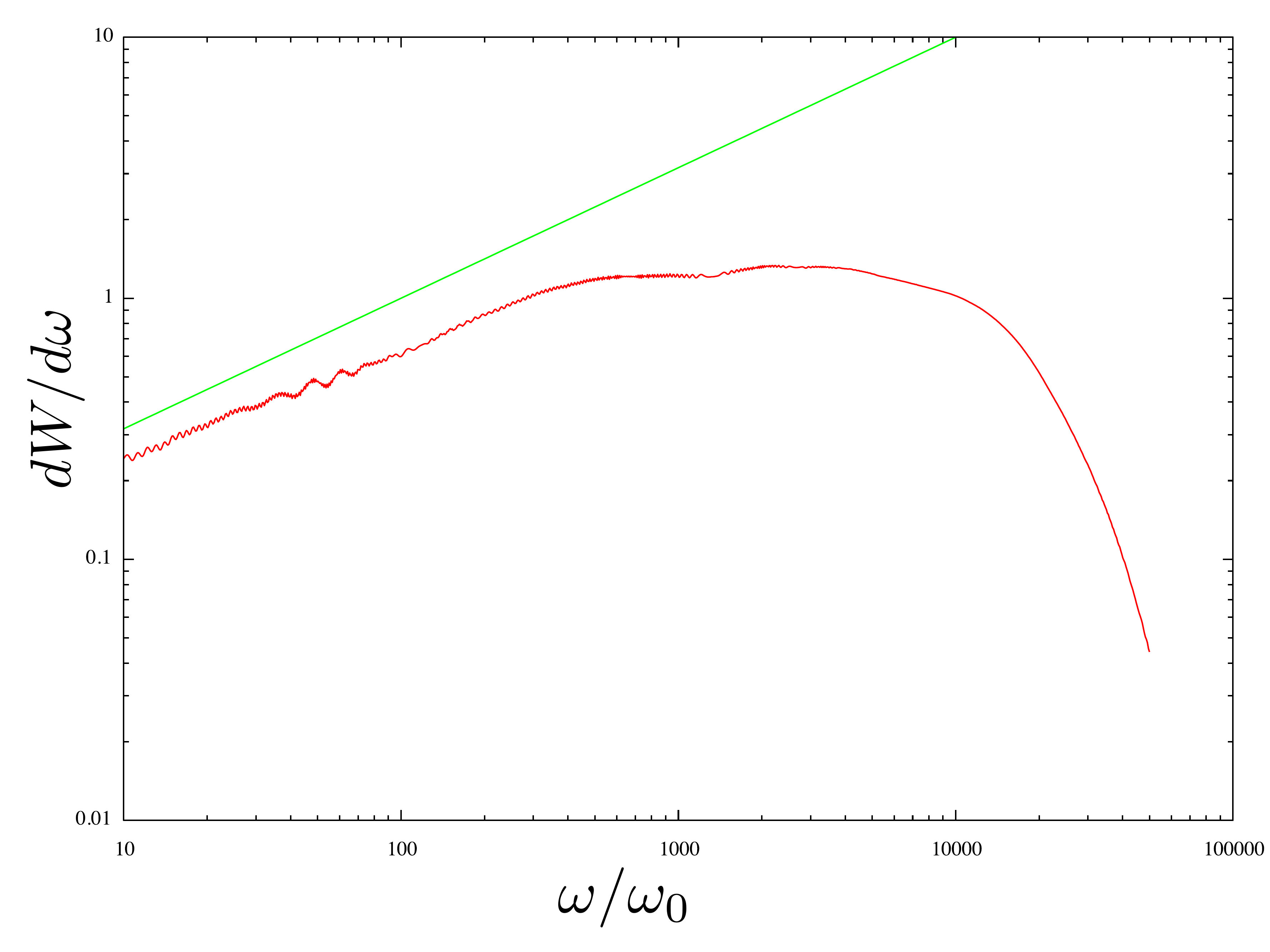}
    }~
    \subfigure[$a=20$]{
      \includegraphics[width=.45\columnwidth]{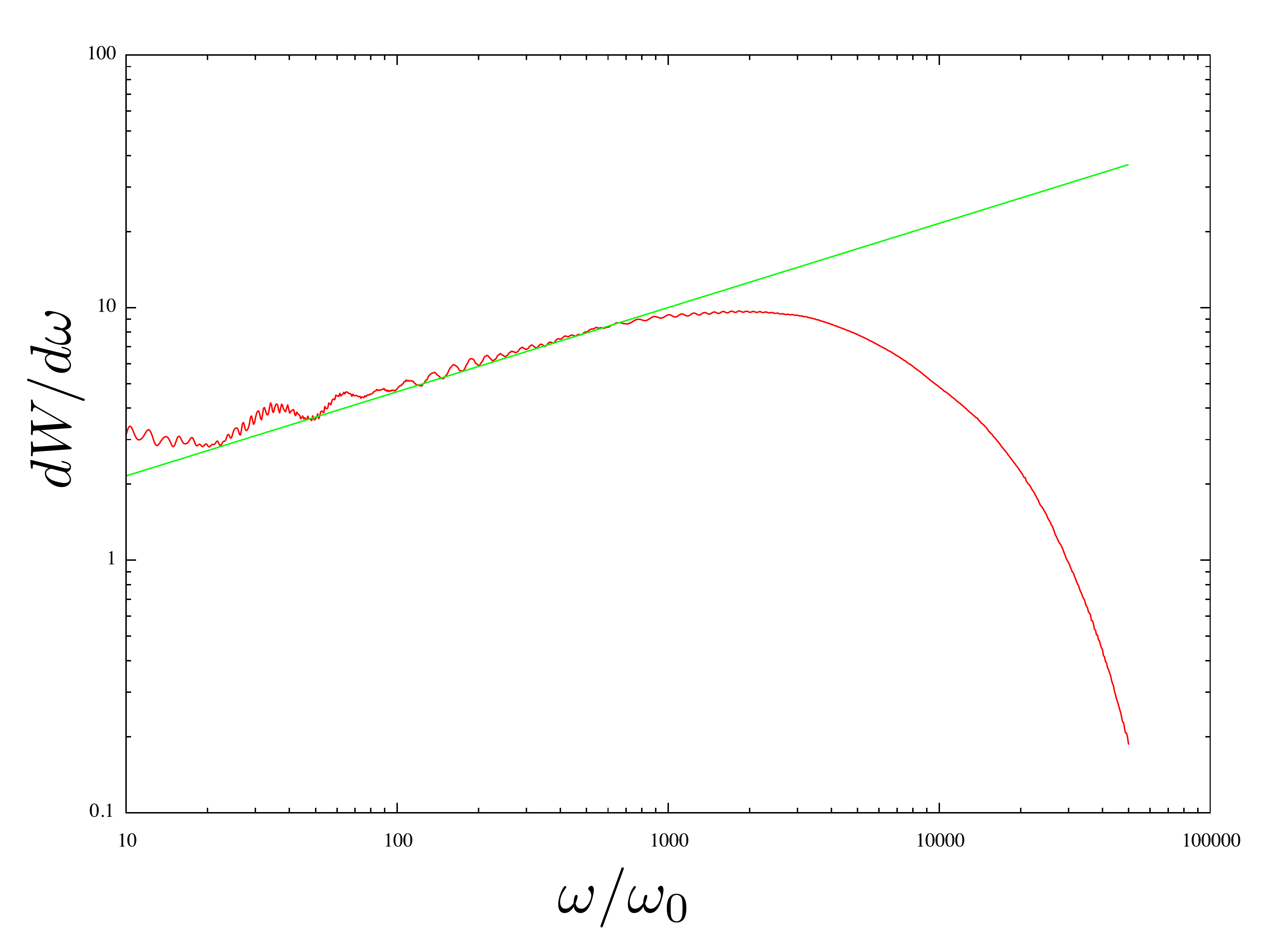}
    }
  \caption{Radiation spectra for $b=\omega_\mathrm{p}/\omega_0=10^{-2}$ and $a=\omega_{\rm st}/\omega_0=0.1$, 
  $5$, $10,$ and $20$.
Vertical axis is the spectral power in arbitrary unit and horizontal axis denotes frequency $\omega$ normalized by $\omega_0$.
The number of electrons used for these calculations is $160$.
(a)~ $a=\omega_\mathrm{st}/\omega_0=0.1$, and the straight line shows a power law spectrum with index $-5/2$:
(b)~ $a=5$, and the straight line shows a power law spectrum with an index $-5/2$:
(c)~ $a=10$, and the straight line shows a power law spectrum with an index $1/2$:
(d)~ $a=20$, and the straight line shows a power law spectrum with index an $1/3$.
We see the transition from jitter radiation regime to the Wiggler radiation regime in Langmuir turbulence (WRL). }
  \label{jit}
  \end{center}
\end{figure}

\begin{figure}
\includegraphics[width=12cm]{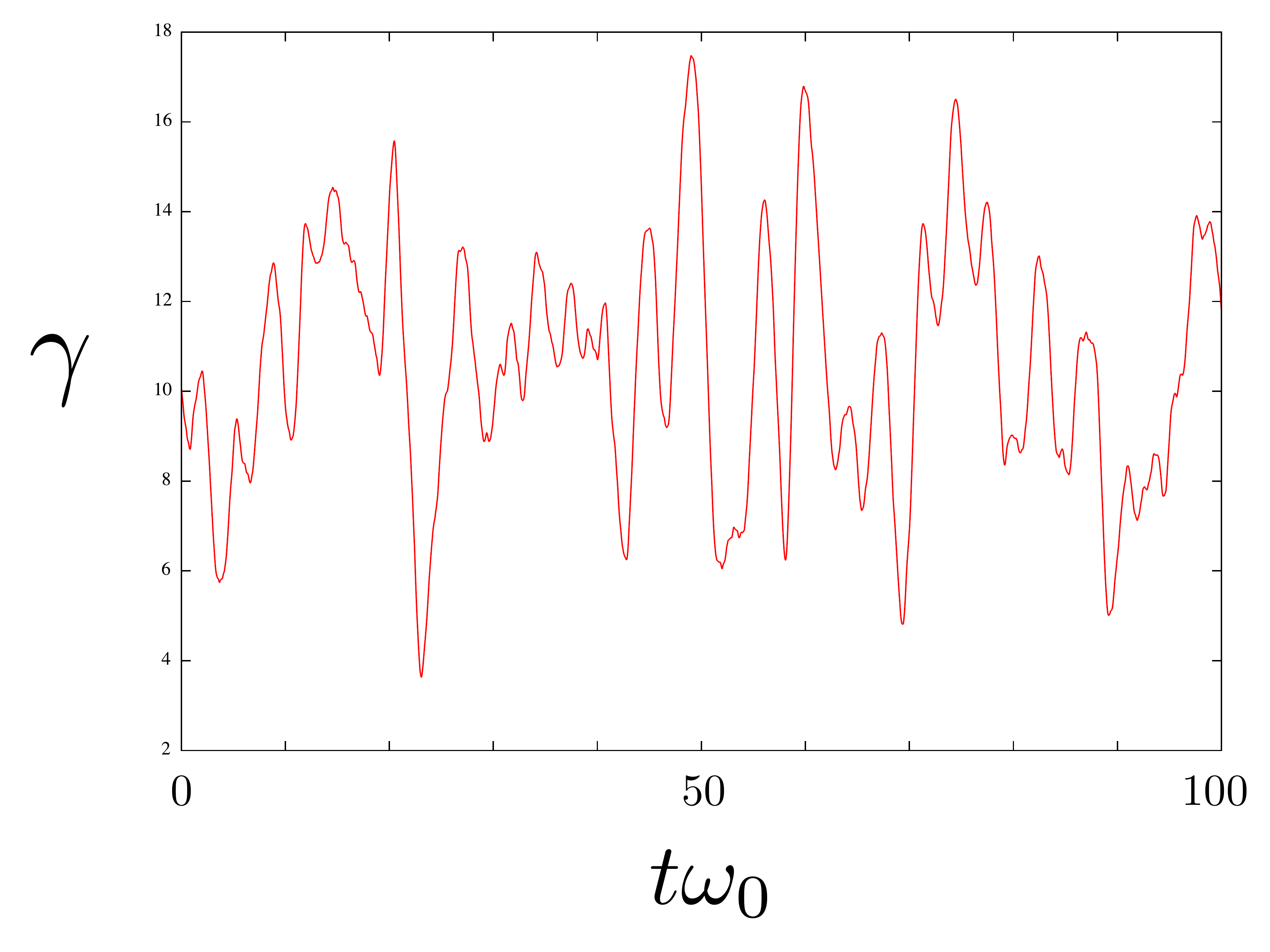}
\caption{An example of the temporal variation of the Lorentz factor of an electron for $a=5$ and $b=10^{-2}$
($\omega_\mathrm{st}=5$, $\omega_0=1$, and $\omega_\mathrm{p}=10^{-2}$).
}
\label{jit_ene}
\end{figure}

First, we show the radiation spectra for $b=\omega_\mathrm{p}/\omega_0\ll1$ (Figure \ref{jit}),
i.e., for the situation where typical spatial scale is shorter than the inertial length.
We set $a= 0.1$ to $20$, and fix $b=10^{-2}$; specifically  we set $\omega_0=1$ and $\omega_\mathrm{p}=10^{-2}$,
and change $\omega_\mathrm{st}$ from $0.1$ to $20$, and take $k_\mathrm{max}=10^3k_\mathrm{min}$.
The inequality $b\ll1$ can be transformed to $\lambda \ll 2\pi c/\omega_\mathrm{p}$, 
which means that the fluctuation scale is much smaller than the inertial length.
These fluctuations would be damped by Landau damping and may not be realized in high energy astrophysical objects (Treumann \& Baumjohan 1997).
However, we cannot reject the possibility that $b\lesssim1$ is realized for some time span in relativistic plasmas, 
so that we study the spectra for $b<1$. To explore the regime for $b<1$ clearly, we set an extreme value $b=10^{-2}\ll1$.
An example of the temporal variation of the Lorentz factor of an electron is depicted in Figure \ref{jit_ene}.

For $a=0.1$,  the low frequency spectrum is as flat as $F_\omega \propto \omega^0$, and there is a break at $\omega\sim 200$, 
and the spectrum declines with power law $F_\omega \propto \omega^{-5/2}$ in the high frequency region.
For $a=5$, the break frequency becomes higher than that for $a=0.1$,
and the high frequency spectrum deviates from a power law.
For $a=10$, the spectrum in low frequency region becomes hard with an index $\sim 1/2$, 
and the spectrum in higher frequencies reveals a cutoff feature.
For $a=20$, we see further different features. 
The spectrum in low frequency side of the peak becomes softer, with the spectral index of about $1/3$,
and we see a small deviation from a power law in the lowest frequency region.
The features for these spectra can be understood by using the analogy to the radiation theory from the stochastic magnetic field
(Medvedev et al. 2011, Teraki \& Takahara 2011).
Since the wavelength of Langmuir waves for $b\ll1$ is very short, 
the oscillation of the electric field can be neglected in the particle crossing time of the wavelength.
In fact, $b\ll1$ is also written by $\lambda/c\ll 2\pi /\omega_\mathrm{p}$.
The crossing time $\lambda/c$ corresponds to the PFT of the typical frequency.

As explained above, we can use the straight analogy to the radiation theory of the stochastic magnetic fields
for the radiation spectra from Langmuir turbulence for $b<1$
by substituting the electric field strength $E$ with magnetic field strength $B$.
For $a<1$ and $b<1$, we can use the radiation theories of the   
DSR and jitter radiation theory (Fleishman 2006, Medvedev 2006).
On the contrary, for $a>1$ and $b<1$ jitter radiation of strong deflection regime 
can be applicable (Medvedev et al. 2011, Teraki \& Takahara 2011).
We call $a<1$ and $b<1$ regime as "jitter radiation" regime because
the jitter radiation is basically perturbative theory for $a<1$.
We call the radiation for $a>1$ and $b<1$ regime as "Wiggler Radiation in Langmuir turbulence", or "WRL" in short.
Although the Wiggler radiation is not the radiation mechanism from the stochastic field but that for 
a fixed field, it has a common picture that the observer is in and off the beaming cone in the course of time.

First, we  describe the signature of the radiation spectra of jitter radiation or Diffusive Synchrotron Radiation.
The radiation signatures are determined by acceleration perpendicular to the motion, and the observer is always in the beaming cone,
which is the same situation as the Langmuir turbulence for $a\lesssim1$ and $b<1$.
For $a\ll1$, the spectrum is written by a broken power law $F_\omega \propto \omega^0$ in the low frequency region 
and $F_\omega \propto \omega^{-\mu+2}$ in the high frequency region. 
The break frequency is $\sim \gamma^2\omega_0=\gamma^2k_\mathrm{min}c$.
For $a\sim 1$, while the break frequency remains the same as $\gamma^2\omega_0$,
the multiple deflection effect comes into play in the spectral features near the break frequency.
The spectrum in the low frequency region becomes $F_\omega \propto \omega^{1/2}$.
Since the multiple deflection makes the angle between observer direction and velocity larger than the beaming cone angle $1/\gamma$,
the observer sees the radiation over the timescale which is determined by the deflection condition.
Fleishman supposed that the angle changes diffusively.
The break frequency of $a\gamma^2\omega_\mathrm{st}$ is calculated from the angle diffusion (Fleishman 2006).
The spectral index of $1/2$ comes from the diffusivity, too.

Next, we describe the signatures for strong deflection regime of $a\gg1$.
For magnetic turbulence, the spectral shape resembles synchrotron radiation 
in the middle frequency region and deviations from it would be seen 
in low frequency region and highest frequency region (Teraki \& Takahra 2011).
The peak frequency of $\gamma^2\omega_\mathrm{st}$ is from the sweeping of the beaming cone on the observer.
The picture is the same as the synchrotron radiation.
For short wavelength regime $\omega_0\gg\omega_{\rm p}$ of Langmuir turbulence, the physical picture is the same as this,
because the spatial fluctuation dominates the changing rate of the deflection angle. 
A single deflection angle is $\sim eE\lambda/\gamma mc^2$,
which is larger than $1/\gamma$ for $a=\omega_\mathrm{st}/\omega_0=e\sigma/mc^2k_\mathrm{min}>1$.
As a result, the beaming cone sweeps out of the observer within one deflection.
The intensity of the radiation off the beaming cone is weak.
Therefore, the timescale which sweeps the observer $\sim 1/\omega_\mathrm{st}$ corresponds to the PFT of the typical frequency.
Considering the photon chasing effect, we get the peak frequency $\sim \gamma^2 \omega_\mathrm{st}$.
The spectral break in the low frequency region would correspond to the deviation from local circular orbit,
but the numerical error from finite integration time is also becoming large in the lower frequency region.
We do not discuss this point further here, since it is not the main point of contents in this paper.
The power law component in the highest frequency region comes from the smaller scale part of turbulence.
It is the same as the spectra of jitter regime.
We note here that the power law component in high frequency region cannot be seen for $a=10$ and $a=20$.
The reason may be as follows.
In contrast to the magnetic turbulence, the energy of radiating electrons changes for $a\gtrsim1$.
Therefore, the high frequency region is determined by only the later part of integration time,
because the peak frequency is $\sim\gamma^2\omega_\mathrm{st}$ for $a>1$, and
the electrons get energy and radiate higher energy photon in later time.
The power law component for the highest frequency region in our calculation is hidden by the component 
that small numbers of electrons with larger energy radiate by Wiggler mechanism.

Finally, we consider effects of the energy change in WRL regime, which is the one of the different points from the magnetic case.
Based on an example of the change of Lorentz factor is depicted in Figure \ref{jit_ene},
we estimate the energy change in PFT $1/\omega_{\rm st}$ for peak frequency.
The PFT is $1/\omega_\mathrm{st}$.
The change of Lorentz factor is estimated as  
\begin{equation}
\Delta\gamma=eE\cdot\frac{c}{\omega_\mathrm{st}}\frac{1}{mc^2}\sim1.
\label{Lfchange}
\end{equation}
Therefore, the change of the Lorentz factor in this timescale is smaller than $1$.
Thus, we do not need to consider the energy change in PFT for peak frequency.
However, for sufficiently large $a$, we have to take it into account for lower frequency spectrum.
We calculate the spectra for modest $a$ in this paper, so we omit this problem.
It will be studied in future works.

\subsection{Long wavelength and weak regime}
\begin{figure}
\includegraphics[width=12cm]{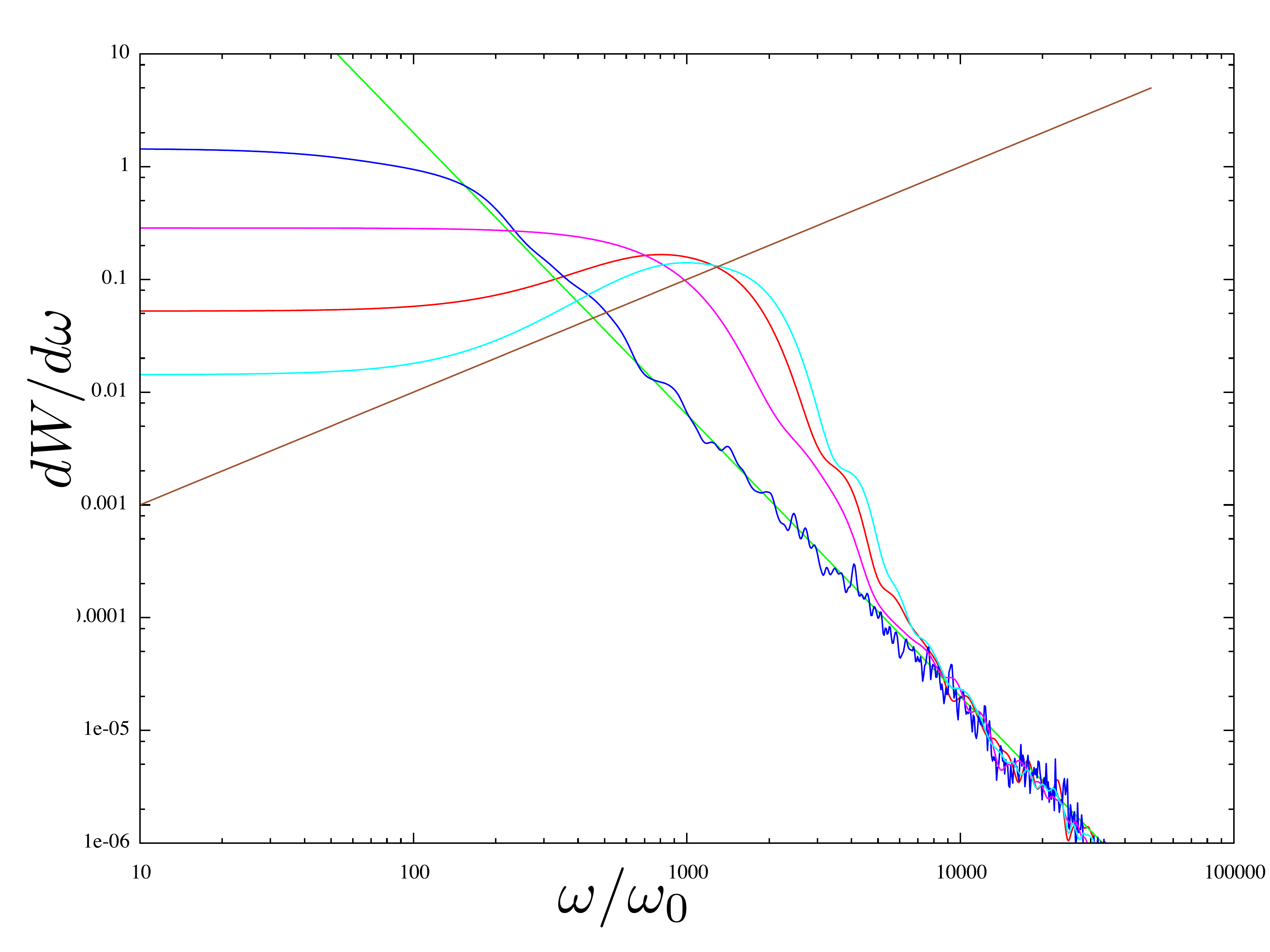}
\caption{Radiation spectra for  $a=10^{-2}$ and $b= 0.1,1, 5, 7$, and $10$ 
from top to down in low frequency range.
The number of electrons used for these calculations is $160$.
The straight lines show a power law spectrum with index $1$ and $-5/2$.
We see the transition from jitter radiation to DRL as $b$ increases.
}
\label{drl}
\end{figure}

\begin{figure}
\includegraphics[width=12cm]{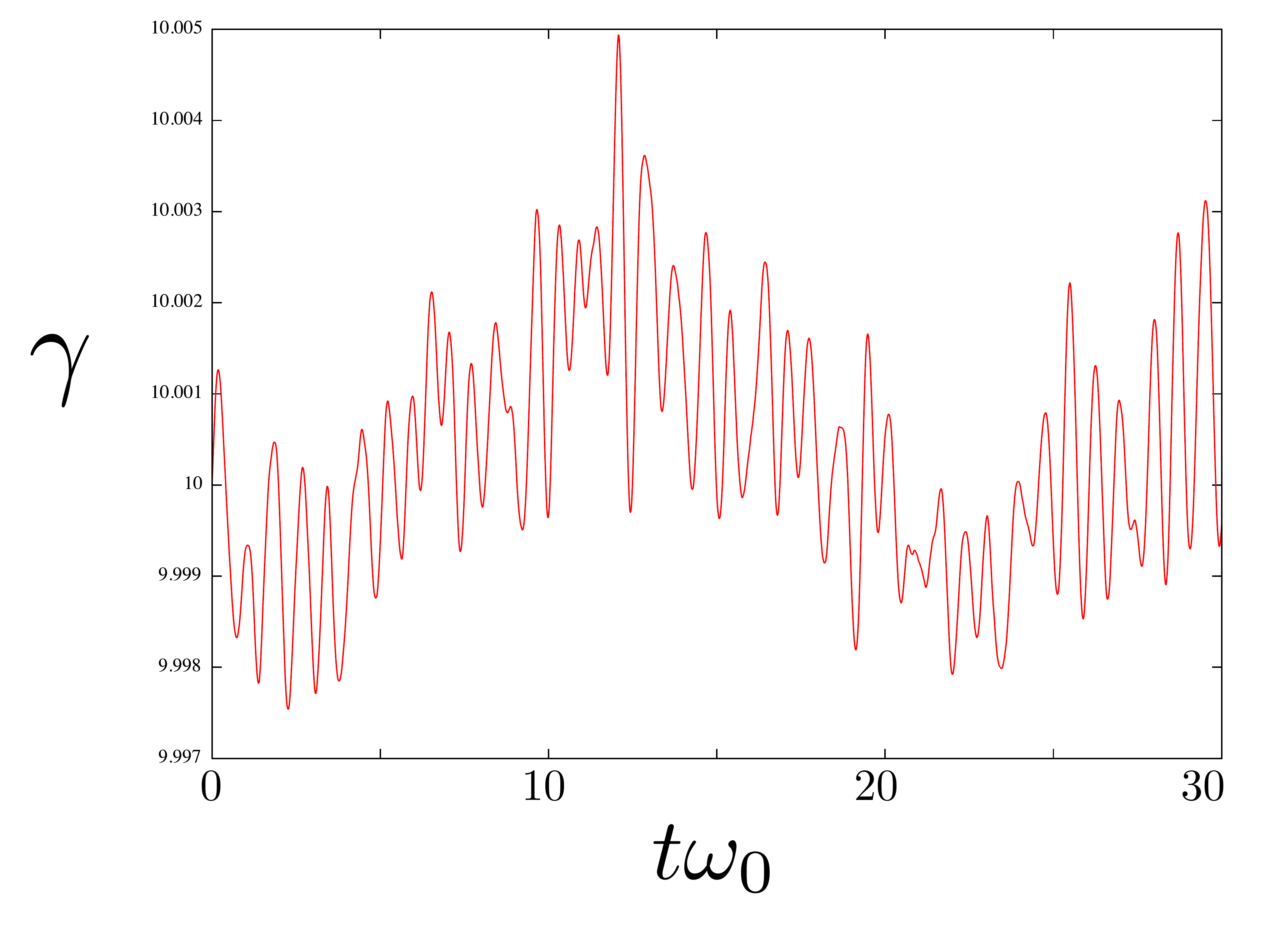}
\caption{An example of the temporal variation of the Lorentz factor of an electron for $a=10^{-2}$ and $b=10$
($\omega_\mathrm{st}=10^{-2}$, $\omega_0=1$, and $\omega_\mathrm{p}=10$).
}
\label{drl_ene}
\end{figure}

Next we show the radiation spectra for $a<1$ and $b>1$ (Figure \ref{drl}), i.e., a situation where long wavelength
turbulence with weak amplitudes dominates; the spectra for $b=0.1$ and $1$ are also depicted for comparison.
An example of the temporal variation of the Lorentz factor of an electron is depicted in Figure \ref{drl_ene}.
Firstly, we consider the meaning of the parameters of $a<1$ and $b>1$, which correspond
to $\omega_\mathrm{p}>\omega_0>\omega_\mathrm{st}$.
In this regime, the changes of the direction of acceleration are mainly due to wave oscillation,
rather than the spatial fluctuations because the crossing time is longer than the oscillation time.
Moreover, $\theta_{\rm def}<1/\gamma$ is derived from $b>a$.
Therefore, we can regard the orbit as straight in the time scale of plasma oscillation $1/\omega_{\rm p}$.
The condition $b>1$ can be transformed to $\lambda>2\pi c/\omega_\mathrm{p}$, 
which means that the inertial length is shorter than the wavelength.
Therefore, this regime is likely to occur in the astrophysical objects.

We set $\omega_0=1$, $\omega_\mathrm{st}=10^{-2}$ and change $\omega_\mathrm{p}$ from $0.1$ to $10$, 
so that $a=10^{-2}$ and $b=0.1$ to $10$, and take $k_\mathrm{max}=10^3k_\mathrm{min}$.
As was discussed in the previous subsection for $b=0.1$, the spectrum shows jitter radiation signature.
The peak frequency is $\gamma^2\omega_0$ and $F_\omega\propto\omega^0$ in the low frequency region, 
and $F_\omega \propto \omega^{-5/2}$ above the peak frequency
reflecting the spectrum of the turbulent electric field $E^2(k)\propto k^{-5/2}$.
As $b$ becomes larger, the peak shifts to higher frequency.
For $b=10$, the peak frequency is $\sim 10^3$, which is identified with $2\gamma^2\omega_\mathrm{p}$.
The spectral index of low frequency side is $\sim 1$.
This feature coincides with the result of DRL theory (Fleishman 2006).
We call this regime "DRL regime".
The DRL theory predicts the spectral shape for $a<1$ and $b>1$ as follows.
The peak frequency is $2\gamma^2\omega_\mathrm{p}$ with an abrupt cutoff in the higher frequency side and 
a power law component emerges in the highest frequency region for $k_\mathrm{max}\gg\omega_\mathrm{p}/c$.
In the low frequency side, the spectrum becomes $F_\omega \propto \omega^1$.

The peak frequency is determined by the timescale of plasma oscillation.
The shortest timescale of the electron motion is $1/\omega_{\rm p}$, and the observer located along the initial 
velocity direction can see this radiation, because the orbit is regarded as straight in this time scale as we showed above.
We consider Doppler boosting, and we get the peak frequency $2\gamma^2 \omega_{\rm p}$ in the observer frame.
The origin of the power law component in the highest frequency region is the same as jitter radiation.
The hard spectral index $1$ in the frequency region lower than the peak is from the photon chasing effect.
The electrons oscillate by the same oscillating frequency $\omega_\mathrm{p}$ and move different angles to the observer
in the observer frame.
The photon chasing effect shifts the photon frequency from each electron and makes the $F_\omega \propto \omega^1$ spectrum
(see Fleishman 2007a,b).
It is regarded as the emission spectrum from an electron integrated over the solid angle,
which can be understood in an analogy to the Undulator theory (Jackson 1999),
although the force direction changes in this case not spatially but temporally.
When the particle mean velocity and the wavenumber are fixed, that makes no difference for the radiation spectra.
The peaky shape of the spectra is the most remarkable feature of the spectra for DRL case.
A difference is that the wave number is a vector, while the frequency is a scalar.
For DSR, the electron feels spatial fluctuation with wavenumber along the velocity $\bm{k}\cdot\bm{v}/v$,
therefore the low frequency spectrum becomes flat.
On the other hand, for DRL, all electrons feel the same frequency of $\omega_\mathrm{p}$ for the Langmuir turbulence
(c.f. Fleishman \& Toptygin 2007a,b and Medvedev 2006).

\subsection{Long wavelength and strong regime}
The remaining interesting regime of $a>1$ and $b>1$ is an open issue.
As we explained in section \ref{int}, the radiation spectrum for Langmuir turbulence 
with $a>b>1$ predicted by Fleishman \& Toptygin (2007b) is $F_\omega \propto \omega^{1/2}$ 
below the peak of $\sim \gamma^2\omega_{\rm p}$,
which is inconsistent with the statements by Getamantsev and Tokarev (1972).
Here we investigate this regime and clarify the radiation spectra.

\begin{figure}
\includegraphics[width=12cm]{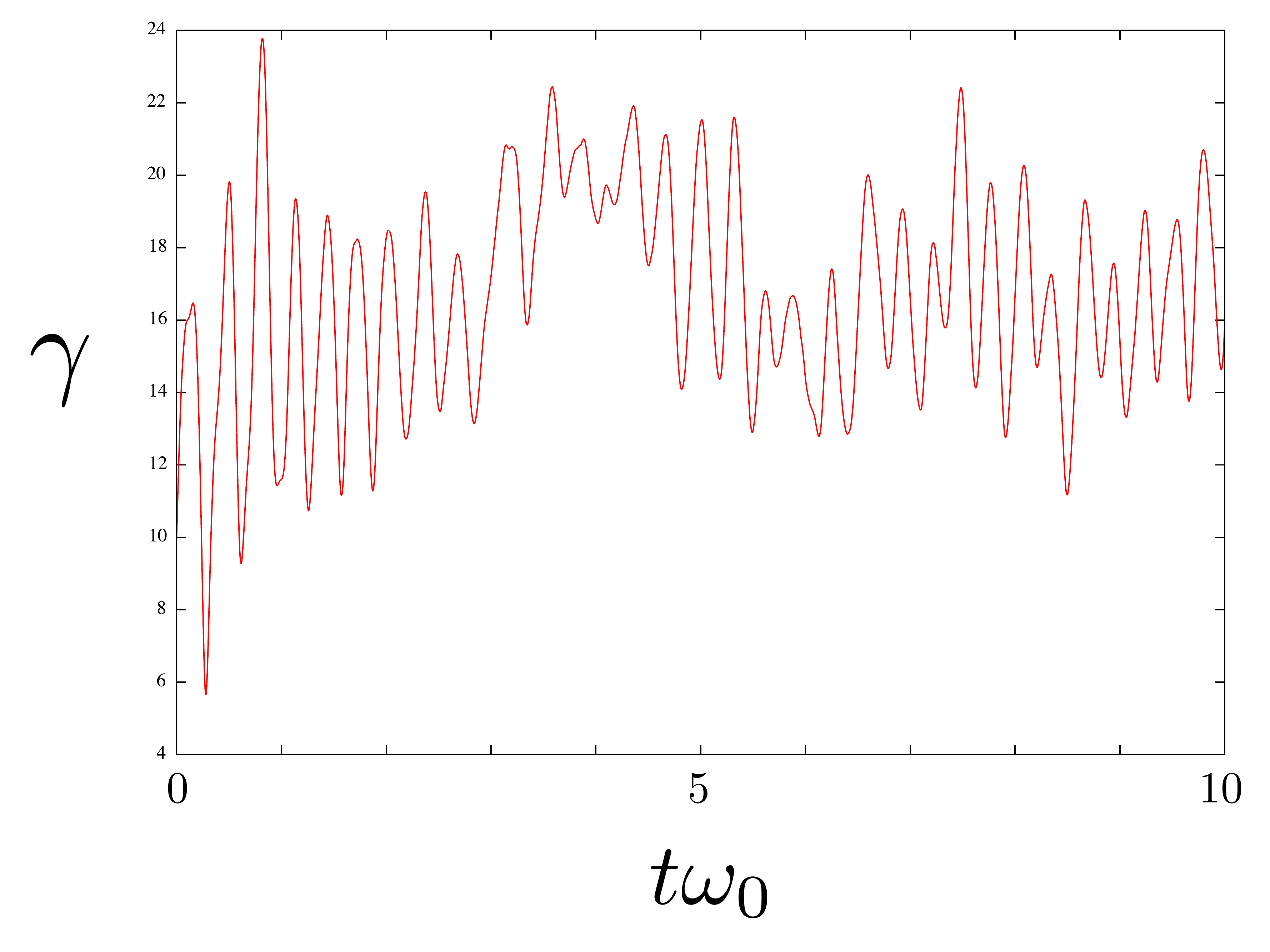}
\caption{An example of the temporal variation of the Lorentz factor of an electron for $a=10^2$ and $b=20$
($\omega_\mathrm{st}=10^2$, $\omega_0=1$, and $\omega_\mathrm{p}=20$).
}
\label{wig_ene}
\end{figure}

The inequations of $a>1$ and $b>1$ mean that the wavelength is longer than the inertial length 
and that the typical scale of PFT is $mc^2/e\sigma$.
We set $a=10^2$, $b=20$ to $800$, so that $a/b=0.125$ to $5$, and $\omega_0=1$.
We set $\omega_\mathrm{st}=10^2$, and we change $\omega_\mathrm{p}$ from $20$ to $800$, 
and $k_\mathrm{max}$ is chosen to be $10k_\mathrm{min}$ here.
We show the interesting results for $a>1$ and $b>1$ (Figure \ref{wig}).
An example of energy change for $a=100$ and $b=20$ is depicted in Figure \ref{wig_ene}.

For clear discussion of the radiation spectra in this regime,
we discuss the radiation spectra for $a>b$ and for $a<b$ separately.
For $a<b$, the peak frequency is $\sim \gamma^2\omega_\mathrm{p}$.
The spectral index of low frequency region is $1$, and cutoff feature can be seen above the peak.
This region is regarded as the DRL regime from these signatures.
For $a<b$, i.e., $\omega_\mathrm{st}<\omega_\mathrm{p}$, the particle is not deflected by large angle because 
the direction of the electric field changes in a time shorter than the time for which the beaming cone sweeps the observer.
We set $k_\mathrm{max}=10k_\mathrm{min}$,  therefore, a power law component in high frequency region is not seen at all.

On the contrary, different features emerge for $a>b$.
The peak frequency becomes larger than $\gamma^2\omega_\mathrm{p}$.
Moreover, the spectra below the peak frequency become softer, the index changes from $1$ to $1/3$.
The energy change of electrons may cause the change of the peak frequency, but it cannot explain the soft spectrum.
Rather, it would be naturally understood that the peak frequency is $\gamma^2\omega_\mathrm{st}$ and 
$F_\omega\propto\omega^{1/3}$ by using the analogy of the Wiggler radiation.
We consider that we should use WRL theory not only for $a>1>b$,
but also for $a>b>1$, because the deflection angle in one deflection is also larger than $1/\gamma$ for this case.
This is in contrast to the DRL theory, which predicts the same spectral signatures
for the parameter range of $a>b>1$ as for $b>a>1$.
Thus, our numerical calculations reveal new features which have not been known previously for Langmuir turbulence.
We ascribe that the parameter regime $a>b>1$ is in the WRL regime in $a-b$ plane.
This radiation feature seems to be consistent with the statements by Getmantsev \& Tokarev (1972).
To clarify the spectral features for $a>b>1$ in more detail, and to confirm their statement and our consideration, 
we examine the radiation from a relativistic electron moving in pure plasma oscillation in the next section.

\section{PURE PLASMA OSCILLATION}
In this section, we investigate the emission of a relativistic electron suffering from pure plasma oscillation in order to 
discuss the interpretation of the features of the radiation spectra for $a>b>1$.
To clarify the origin of the peak frequency $\gamma^2\omega_{\rm st}$, we set a simple configuration of the electric field,
where electron motion is deterministic compared to stochastic character in turbulent fields.
We calculate the electron velocity analytically and the radiation spectra numerically. 
By comparing the motion and spectra, we interpret the mechanism which determines the peak frequency.
Lastly we consider the radiation spectra from the turbulent field by using these results.

We use a single Langmuir wave which has infinitely large wavelength $k=0$, in other words, $\omega_0=0$. 
Therefore, it is a pure plasma oscillation.
We set $\bm{E}=(E_x,0,0)$ with
\begin{equation}
E_x=E_0 \cos(\omega_\mathrm{p}t).
\end{equation}
We characterize the field by using a single parameter of 
\begin{equation}
  \eta \equiv \omega_\mathrm{st}/\omega_\mathrm{p}.
\end{equation}
We inject an electron along the $z$-axis at $t=0$ with the initial Lorentz factor $\gamma_\mathrm{init}$
and solve the equation of motion (\ref{eom}).
Therefore, the orbit is determined by $\gamma_\mathrm{init}$ and $\eta$.
Solving the equation of motion, we get the momentum
\begin{equation}
  \begin{aligned}
    p_x =&& - \frac{eE_0}{\omega_\mathrm{p}}\sin{\omega_\mathrm{p}t}\\
    p_z =&& \gamma_{\rm init}\beta_{\rm init}mc = && \mathrm{const}.
  \end{aligned}
\end{equation}
Since the velocity is a periodic function, we can define the mean velocity by
$\bar{\beta}=\frac{\omega_\mathrm{p}}{2\pi}\int^{2\pi/\omega_\mathrm{p}}_0\beta_zdz$.
The mean velocity $\bar{\beta}$ cannot be represented elementarily in a general form.
Then, we take the parameter $\eta\ll\gamma_{\rm init}$ and approximate the motion hereafter.
We expand the Lorentz factor and the velocity,
and get the mean velocity and the mean Lorentz factor in the lowest order of $\eta/\gamma_{\rm init}$.
\begin{equation}
  \begin{aligned}
    \bar{\beta}=&&\beta_\mathrm{init}(1-\frac{\eta^2}{4\gamma_{\rm init}^2})\\
    \bar{\gamma}=&&\frac{\gamma_\mathrm{init}}{\sqrt{1+\frac{\eta^2}{2}}}
  \end{aligned}
\end{equation}
For $\eta\ll1$, $\bar{\gamma}\sim \gamma_\mathrm{init}$, while for $\eta\gg1$,
$\bar{\gamma}=\sqrt{2}\gamma_{\rm init}/\eta$.
We note that $\bar{\gamma}$ can be much smaller than $\gamma_\mathrm{init}$ for $\eta\gg1$.
Using this approximated velocity, we calculate the maximum Lorentz factor in the mean velocity frame, 
to clarify the fact that the radiation signatures depend on $\eta$ as
\begin{equation}
 \gamma'_\mathrm{max}=\frac{\gamma_\mathrm{init}}{\sqrt{1+\frac{\eta^2}{2}}}
                      \left[\sqrt{\gamma_\mathrm{init}^2+\eta^2} - 
                      \beta_\mathrm{init}(1-\frac{\eta^2}{4\gamma_{\rm init}^2})\sqrt{\gamma_\mathrm{init}^2-1}\right].
\end{equation}
For $\eta\ll1$, 
\begin{equation}
\gamma_\mathrm{max}' = 1+\frac{\eta^2}{2}.
\end{equation}
The motion in this frame is non-relativistic, therefore, the radiation in this frame is dipole radiation.
On the other hand, for $\eta\gg1$, the maximum Lorentz factor is 
\begin{equation}
\gamma_\mathrm{max}' = \frac{3\sqrt{2}}{4}\eta.
\end{equation}
Therefore, the motion is relativistic even in this frame and
the radiation spectrum consists of higher harmonics, because $\beta'$ approaches $1$.
It should be noted that for $\eta=1$, the motion in the mean velocity frame is 
mildly relativistic with Lorentz factor $\gamma'_\mathrm{max}=1.02$, and $\beta'=1/5$.
We can see that the transition from non-relativistic harmonic motion to relativistic motion
occurs around $\eta\sim$ a few from this fact.

\begin{figure}[htb]
  \begin{center}
    \subfigure[$\eta=10^{-3}$]{
      \includegraphics[width=.45\columnwidth]{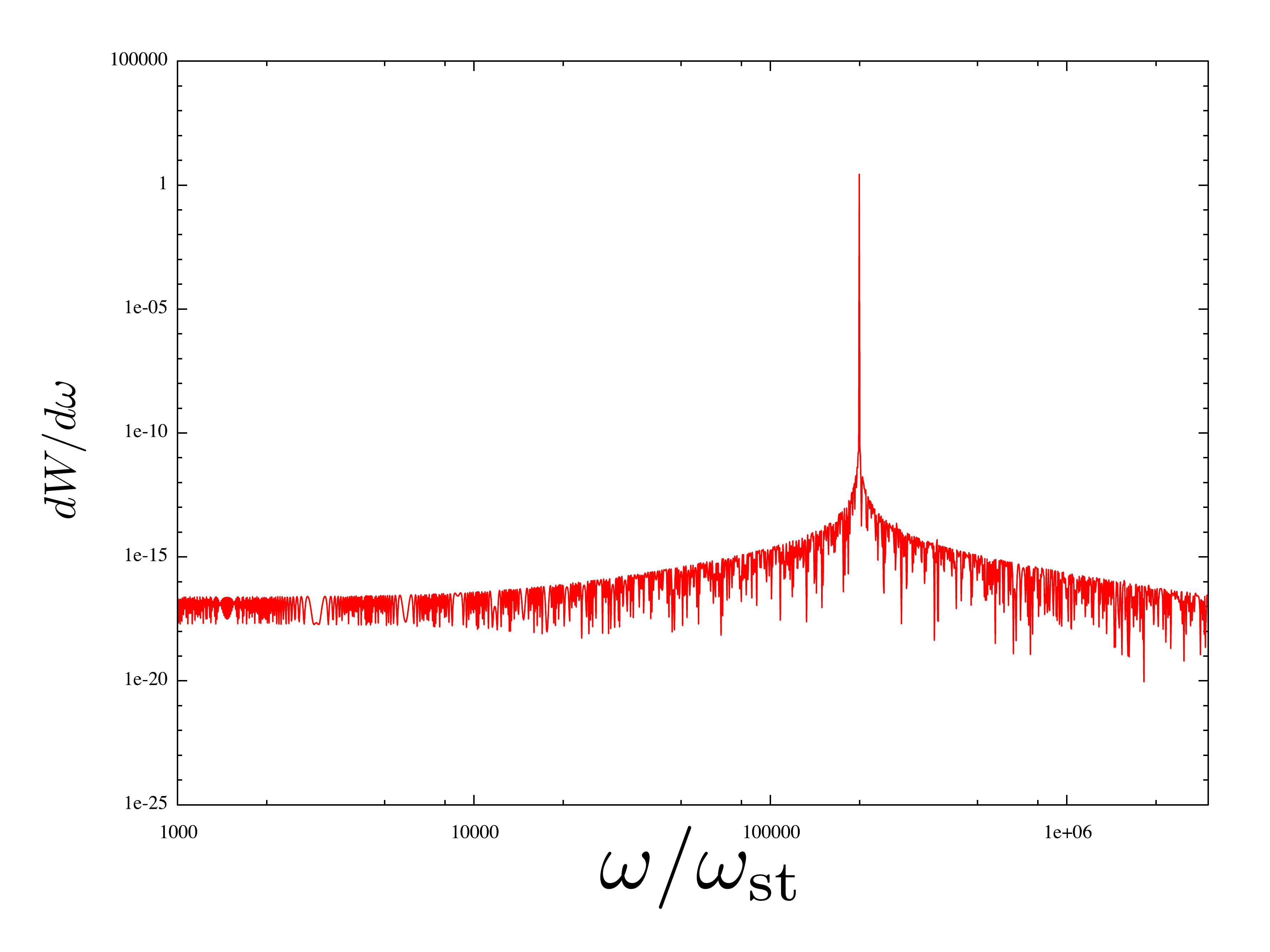}   
    }~
    \subfigure[$\eta=10^{-2}$]{
      \includegraphics[width=.45\columnwidth]{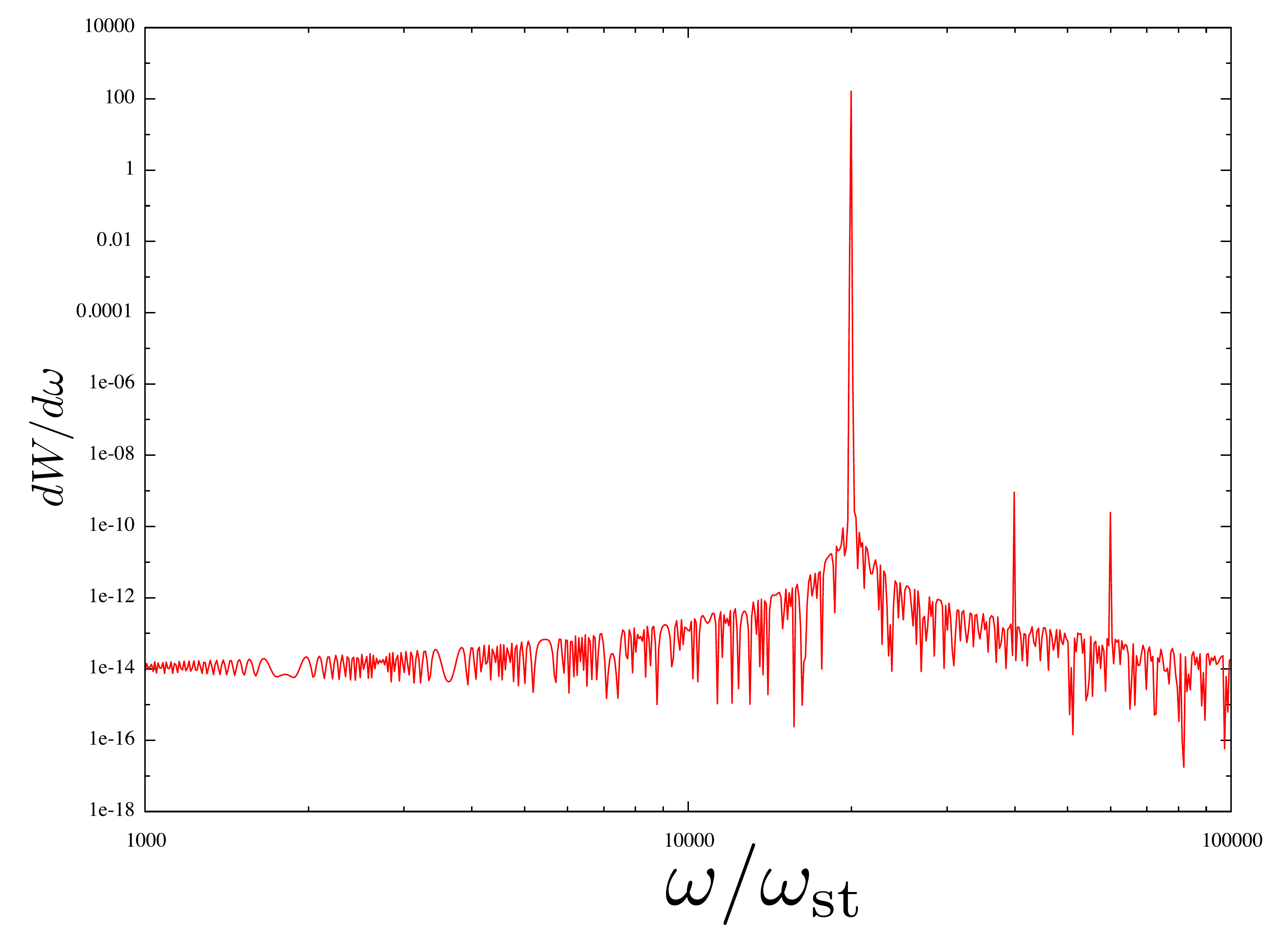}
    }\\
    \subfigure[$\eta=1$]{
      \includegraphics[width=.45\columnwidth]{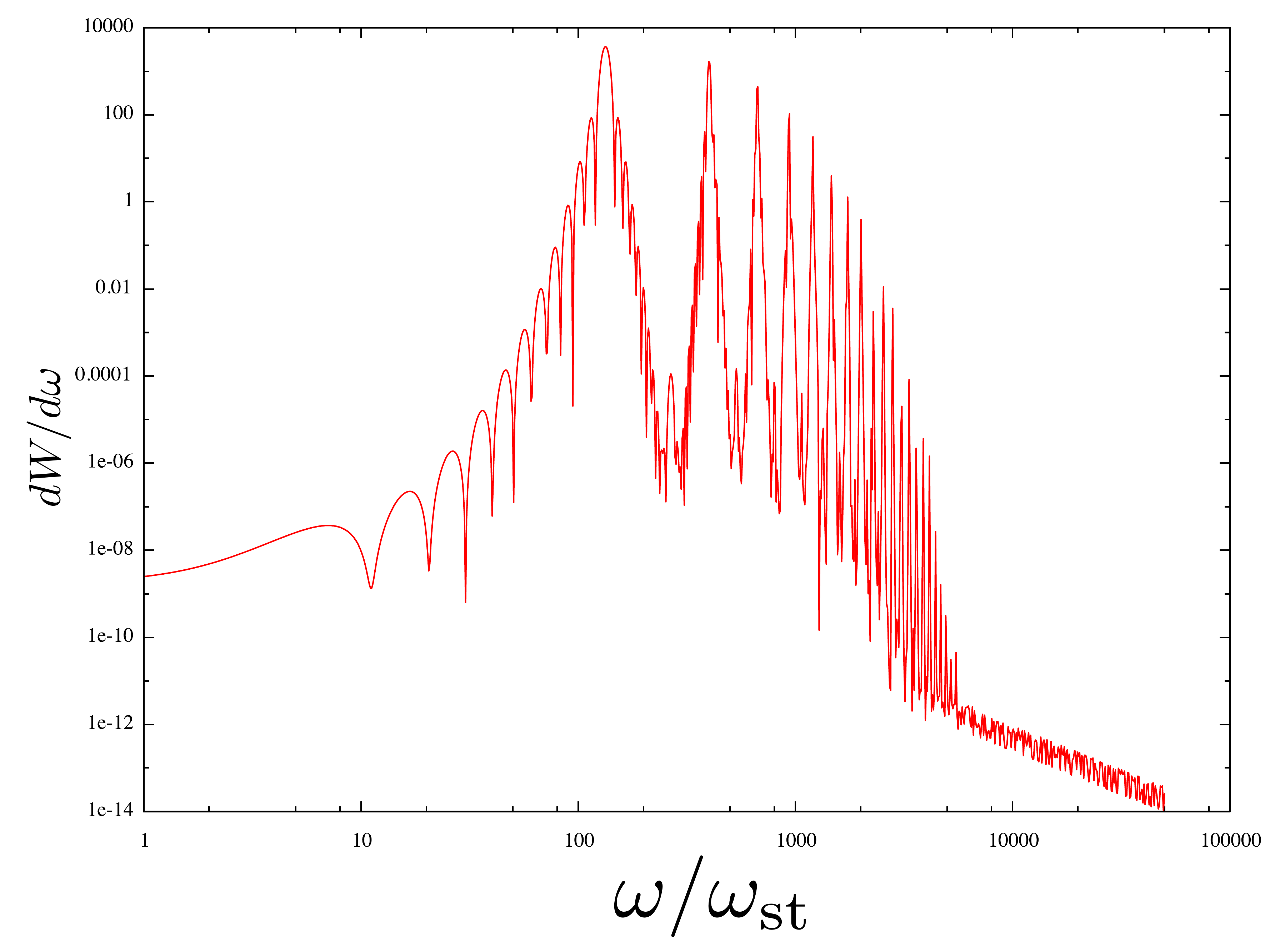}
    }~
    \subfigure[$\eta=500$]{
      \includegraphics[width=.45\columnwidth]{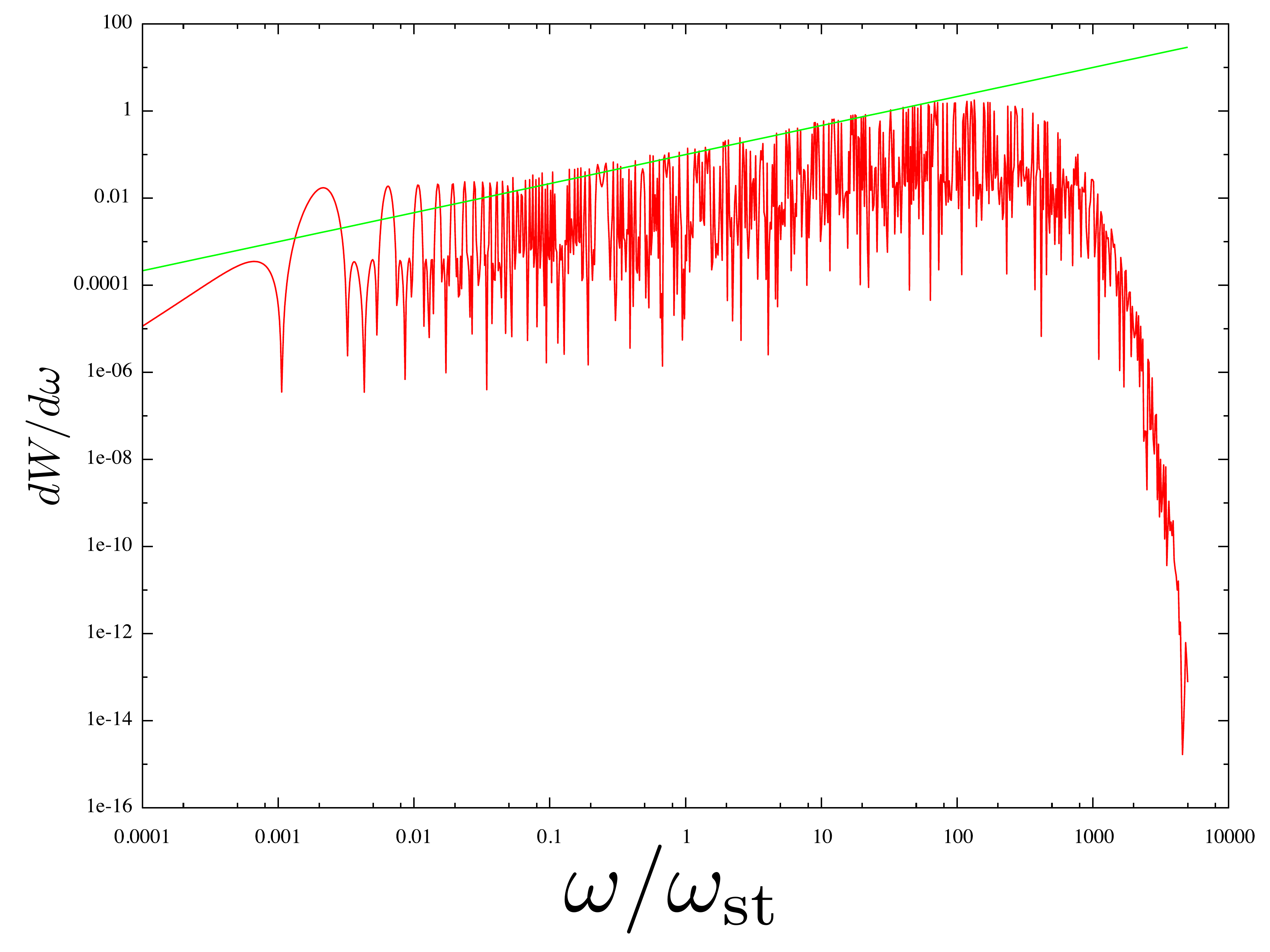}
    }
  \caption{Radiation spectrum from an electron with $\gamma_\mathrm{init}=10$ moving in the oscillating field.
Horizontal axis is frequency normalized by $\omega_\mathrm{st}=1$.
(a)~ $\eta=\omega_{\rm st}/\omega_{\rm p} = 10^{-3}$, ($\omega_\mathrm{p}=10^3$)
(b)~ $\eta=10^{-2}$, ($\omega_\mathrm{p}=10^2$)
(c)~ $\eta=1$, ($\omega_\mathrm{p}=1$)
(d)~ $\eta=500$, ($\omega_\mathrm{p}=2\times 10^{-3}$), and the straight line shows a power law spectrum with index $2/3$.
The fundamental mode in panels (a) and (b) are $\bar{\gamma}^2\omega_{\rm p}\simeq \gamma_{\rm init}^2\omega_{\rm p}$,
while that in (c) is $\bar{\gamma}^2\omega_{\rm p}=133<\gamma_{\rm init}^2\omega_{\rm st}$.
We see the strong higher harmonics in (c) and (d).
  }
  \label{peaks}
  \end{center}
\end{figure}

Next we show numerically calculated radiation spectra from the electron 
and their features are interpreted in terms of the properties of the orbit.
We fix $\omega_\mathrm{st}=1$, and change $\omega_\mathrm{p}$ to change the parameter $\eta$.
The observer is on the $z$-direction.
We calculate radiation spectra using much longer integration time than the PFT, because the electron moves perfectly periodically.
As a consequence, the spectra show very sharp features, which makes it easier to understand the relation between spectral 
features and orbit.
First, we show the spectrum for $\eta= 10^{-3}$ ($\omega_\mathrm{p}=10^3$, Figure \ref{peaks}(a)).
We see a sharp peak like a delta function at the frequency $2\gamma_\mathrm{init}^2 \omega_\mathrm{p} = 2\times10^5$.
This is understood in terms of the motion of the electron in the mean velocity frame.
For $\eta\ll1$, in the mean velocity frame, electron motion is a non-relativistic simple harmonic motion 
with the frequency $\sim \bar{\gamma}\omega_\mathrm{p}$.
Therefore, the radiation is the dipole radiation with the frequency of $\bar{\gamma}\omega_\mathrm{p}$.
Since $\bar{\gamma}\sim\gamma_\mathrm{init}$ for $\eta\ll1$, 
the radiation frequency in the observer frame is $2\gamma_\mathrm{init}^2\omega_\mathrm{p}$.
Thus, we ascribe the frequency $2\gamma_\mathrm{init}^2\omega_\mathrm{p}$
in the radiation spectra of perturbative regime ($\omega_\mathrm{st}<\omega_\mathrm{p}$) to the Doppler shifted dipole radiation.
Next we show the spectrum for $\eta = 0.01$ ($\omega_\mathrm{p}=10^2$, Figure \ref{peaks}(b)).
We can see the higher harmonics of $2\gamma_\mathrm{init}^2 \omega_\mathrm{p}=2\times10^4$.
It is from effects of retarded time, as is clearly seen in the mean velocity frame.
However, the effect is very weak, as the power of the second harmonics is about $10^{11}$ 
times smaller than the fundamental mode in the frequency resolution of this calculation.
The ratio of the power of the second harmonics to the fundamental mode is proportional to $\beta'^2$,
so the second harmonics is much smaller than the fundamental mode in this case.

\begin{figure}
\includegraphics[width=12cm]{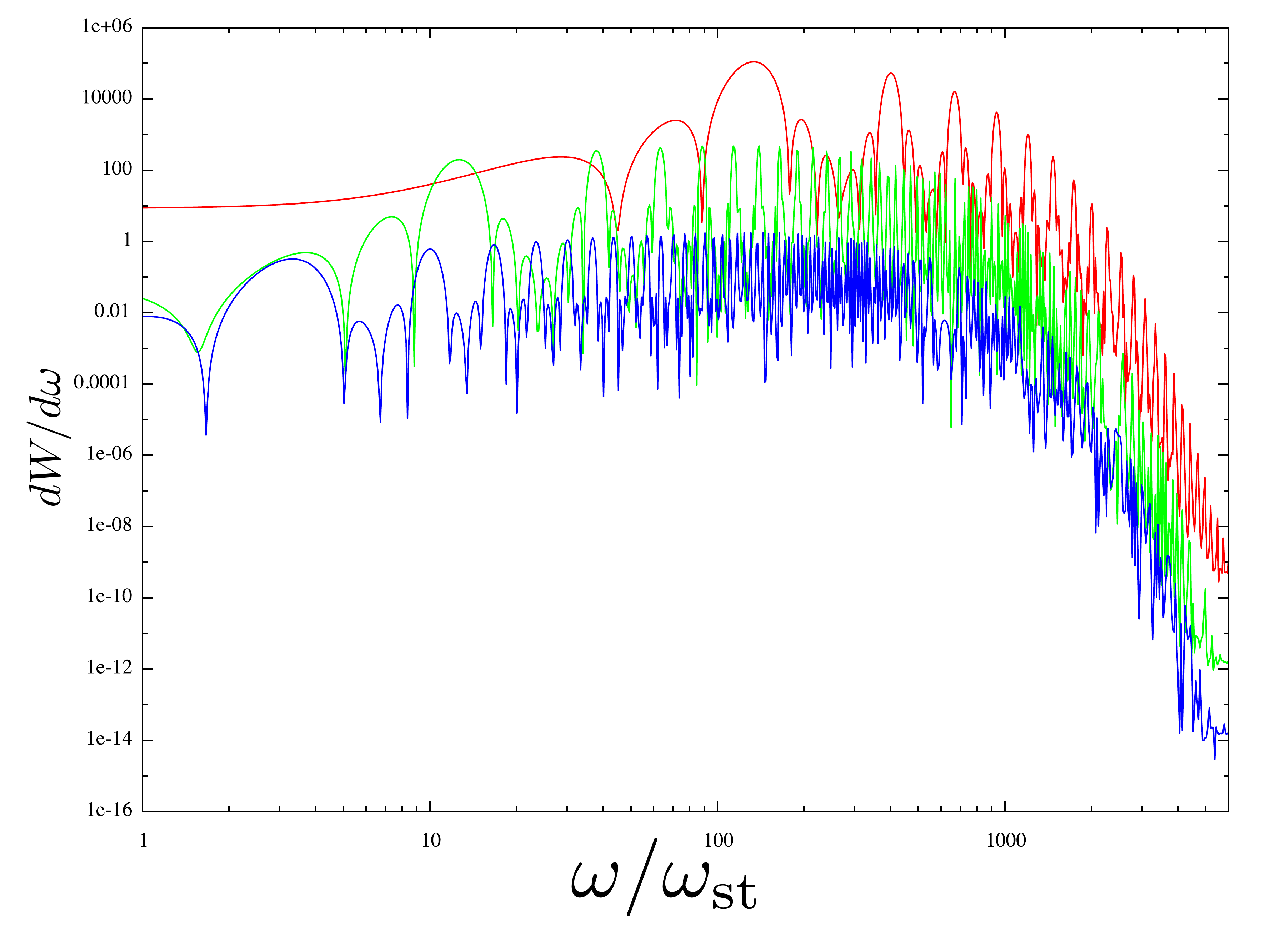}
\caption{Radiation spectra for $\eta=1, 3$, and $5$ ($\omega_\mathrm{p}=1,1/3$, and $1/5$) from top to down.
A factor of $10^{2}$ is multiplied to the spectrum for $\eta=1$, 
and $10^{-2}$ is multiplied for $\eta=5$.
The fundamental frequency is $2\bar{\gamma}^2\omega_\mathrm{p}$. 
It is $133$ for $\eta=1$, $12$ for $\eta=3$ and $3$ for $\eta=5$.
The cutoff frequency is a few times of $\gamma_{\rm init}^2\omega_{\rm st}=100$,
which does not depend on $\omega_{\rm p}$ explicitly.
}
\label{nc_peak}
\end{figure}

For $\eta\sim1$, the spectral shape changes significantly.
First, the higher harmonics stand more strongly, because $\beta'$ approaches $1$.
Many harmonics are as strong as the fundamental mode for $\eta =1$ (Figure \ref{peaks}(c)),
and the envelope of the peaks of the harmonics shows an exponential cutoff. 
We note that the spectrum in the frequency region higher than $5000$ comes from numerical error.
Second, the frequency of the fundamental mode deviates from $2\gamma_\mathrm{init}^2\omega_\mathrm{p}$,
because the difference between $\bar{\gamma}$ and $\gamma_\mathrm{init}$ becomes larger.
The mean Lorentz factor $\bar{\gamma}$ is $\sqrt{2/3}\gamma_\mathrm{init}$ for $\eta=1$, thus 
the frequency of the fundamental mode in the original frame is $2\bar{\gamma}^2\omega_\mathrm{p}=133$.
The difference between $133$ and $2\gamma_\mathrm{init}^2\omega_\mathrm{p}=200$ is small,
but we can discern it in Figure \ref{peaks}(c).
Next we discuss the peak frequency (cutoff frequency) for $\eta>1$.
We show the spectra for $\eta=1,3$, and $5$ in Figure \ref{nc_peak}.
The fundamental frequency is $133$ for $\eta=1$, $12$ for $\eta=3$, and $3$ for $\eta=5$, 
but the cutoff frequency around a few hundreds does not change.
We see the cutoff frequency is not of the fundamental mode, but it is determined by the higher harmonics for $\eta>1$.
The radiation spectra in the observer frame also can be derived by regarding it as a Doppler boosted emission.
However, since the mechanism which determines the peak frequency is the same as the Wiggler radiation
we can understand the peak frequency more easily by considering the PFT in the observer frame.
The condition $\eta=\omega_\mathrm{st}/\omega_\mathrm{p}>1$ is equivalent to that 
PFT of the typical frequency in Wiggler mechanism, where $1/\omega_\mathrm{st}$ is
shorter than the oscillating time $1/\omega_\mathrm{p}$.
On the other hand, the Lorentz factor relevant for the peak frequency is not $\bar{\gamma}$, but $\gamma_\mathrm{init}$,
because the beaming cone sweeps the observer in the phase around $2n\pi$, where $n$ is a natural number.
We note that the change of the Lorentz factor in the time scale of $1/\omega_\mathrm{st}$ is $1$ at most, 
as seen in equation (\ref{Lfchange}).
As a result, the cutoff frequency is $\sim \gamma_\mathrm{init}^2\omega_\mathrm{st}$.
In this way, we get a clear understanding of the mechanism of the peak frequency shift around $\eta \sim O(1)$.

\begin{figure}
\includegraphics[width=12cm]{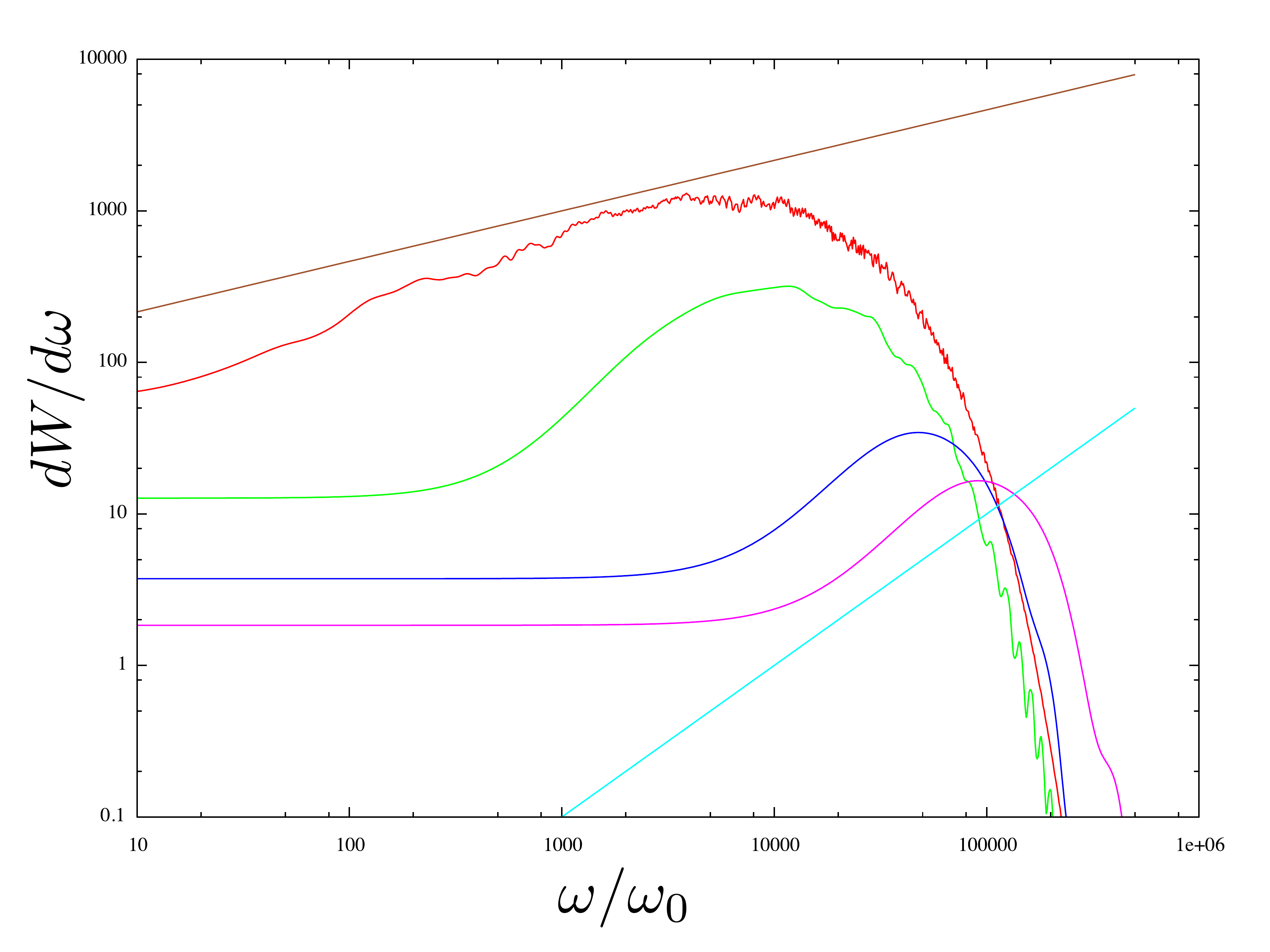}
\caption{Radiation spectra for $a=10^2$ with $b=20$, $90$, $400$, and $800$ from top to down.
The number of electrons used in these calculations is $800$.
The straight lines show the power law spectra with index $1/3$ and $1$.
We see the transition from the DRL to WRL as $b$ decreases. 
}
\label{wig}
\end{figure}

Here, we compare the results in this section with the radiation spectra obtained in the preceding section.
The case of Langmuir turbulence with $a>b>1$ ($\omega_\mathrm{st}>\omega_\mathrm{p}>\omega_0$)
corresponds to the case of pure plasma oscillation with $\eta>1$,
since $\eta=\omega_\mathrm{st}/\omega_\mathrm{p}$ and $\omega_0=0$ for pure plasma oscillation.
Moreover, the approximation we used in pure plasma oscillation of  
$\gamma_\mathrm{init}>\eta$ is also applicable for the case of Langmuir turbulence, 
since $\eta=a/b\leq5$ and $\gamma_\mathrm{init}=10$ for the spectra shown in Figure \ref{wig}.
For pure plasma oscillation with $\eta>1$, we have shown that the peak frequency is $\gamma_\mathrm{init}^2\omega_\mathrm{st}$,
and it consists of the higher harmonics of $\bar{\gamma}^2\omega_\mathrm{p}$.
Therefore, the peak frequency for the Langmuir turbulence with $a/b>1$ in Figure \ref{wig} is interpreted as $\gamma^2\omega_\mathrm{st}$,
and it naturally explains the fact that the peak frequency is larger than $\gamma^2\omega_\mathrm{p}$.
Lastly we consider the spectral index for the Langmuir turbulence with $a>b>1$.
As we see above, the spectral index for $a>b>1$ is neither $1$ predicted by perturbative DRL 
nor $1/2$ predicted by the angle diffusion effect.
We regard that the spectral index is around $1/3$,
because the radiation mechanism is identified as Wiggler mechanism, and the angle integrated spectral index is $1/3$ in Wiggler mechanism. 
The energy change in PFT for the frequency lower than peak can be larger than $1$ for $a>b>1$.
This is a different point from the Wiggler, and it can affect forming the spectral index.
However, we choose modest value of $a/b$ for Langmuir turbulence in the present paper, so that this effect does not stand out.
Summarizing above, we confirm that the peak frequency is $\sim \gamma^2\omega_\mathrm{st}$ 
and the spectral index in low frequency side of the peak $\sim 1/3$ for the Langmuir turbulence with $a>b>1$ and $a/b=O(1)$.
The mechanism which determine these radiation features is Wiggler mechanism, which is consistent with the statements
by Getamantsev and Tokarev (1972).

\begin{figure}
\includegraphics[width=12cm]{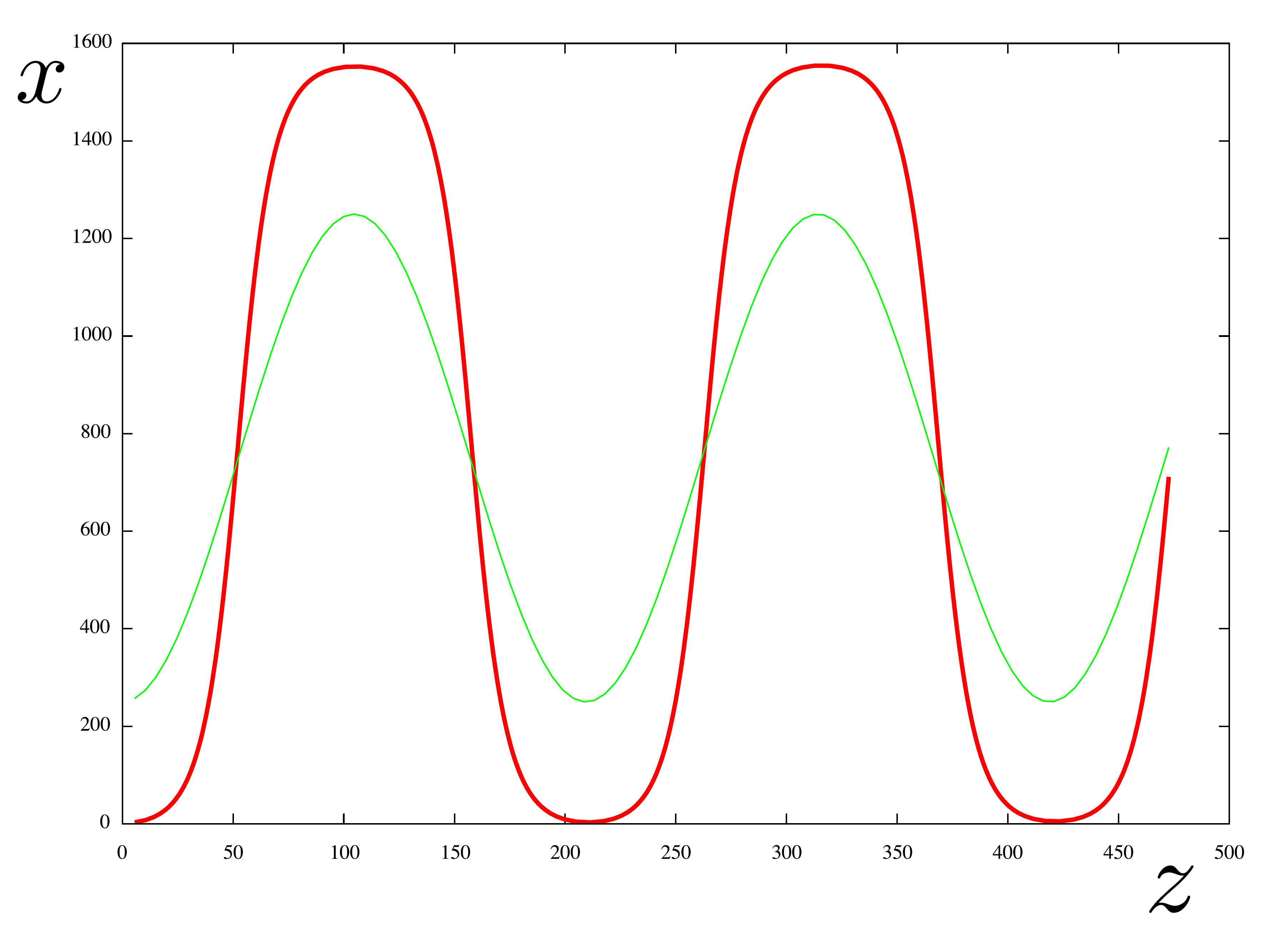}
\caption{Electron orbit for $\eta =500$.
Vertical axis is $x$, and horizontal axis is $z$.
The thick line shows the orbit for the radiating electron,
while the thin line shows a sine curve for comparison.
}
\label{tr500}
\end{figure}

Lastly we show an example of the extreme case of $\eta\gg\gamma_\mathrm{init}$.
For $\eta\gg \gamma_\mathrm{init}$, the motion becomes strongly nonlinear and cannot be treated analytically.
Thus, we show numerically calculated electron orbit.
We show the radiation spectra and the orbit for $\eta=500>\gamma_\mathrm{init}$ (Figure \ref{peaks}(d), Figure \ref{tr500}).
As we expected, the peak frequency is $\sim \gamma_\mathrm{init}^2\omega_\mathrm{st}$,
because the observer sees this electron mainly in the phase that the electric field and the electron velocity
is nearly perpendicular.
We note that the spectral index of $2/3$ is the same as the Wiggler radiation 
when the observers located in a particular direction, i.e., the spectrum is not the angle integrated spectrum.
This spectral index is an collateral evidence for the spectral index of $1/3$ for the case of turbulence with $a>b>1$
for which angle integrated spectrum is calculated.
The Lorentz factor changes from $10$ to $O(100)$, 
but $\gamma_\mathrm{init}$ is relevant to the observer situated at $z$-axis.
In other words, the beaming cone sweeps the observer when $\gamma \sim \gamma_\mathrm{init}$
in the present geometry.
The Lorentz factor relevant to an observer oriented in different angle is significantly different to each other.
Moreover, in general the energy change in the PFT of peak frequency becomes larger than $mc^2$,
because the electric force in some phase of oscillation accelerates the electron linearly,
and the curvature radius becomes larger.
We have to consider the linear acceleration emission in this case.
Thus, we draw the line on $a/b=\gamma$ and $a=1$, and divide the $a>b>1$ region.
We call the radiation for this parameter range as "non-linear trajectory" radiation.
This part of the spectra from an electron moving in the 3D turbulent electric field is determined by the chaotic trajectory.
The generalization of the features of this regime is a future work.

Summarizing this section, we have considered the motion and radiation in a single mode plasma oscillation.
We clarify that the cutoff frequency for $\eta>1$ ($\omega_\mathrm{st}>\omega_\mathrm{p}$) is $\gamma_\mathrm{init}^2\omega_\mathrm{st}$,
which consists of higher harmonics of the fundamental frequency of $\bar{\gamma}^2\omega_\mathrm{p}$.
It is from the effect that the beaming cone sweeps the observer, in the same way as the Wiggler radiation. 
Using this result, we interpret that the peak frequency for 3D Langmuir turbulence for $a>b>1$ 
($\omega_\mathrm{st}>\omega_\mathrm{p}>\omega_0$) is $\gamma^2\omega_\mathrm{st}$, where $\gamma$ is determined by the acceleration.
The shallower spectrum for $a>b>1$ can be explained by WRL mechanism.
Lastly, we show numerically calculated radiation spectra for the extreme case of $\eta\gg\gamma_\mathrm{init}$.
It shows Wiggler like spectra in the middle and high frequency region, while the flattening can be seen in the low frequency region.
It is from the effect of elongated trajectory to the electric field direction.
The radiation signatures are summarized as a chart in the $a-b$ plane in Figure \ref{a-b_plane}. 

\section{DISSCUSSION \& SUMMARY}
\begin{figure}
\includegraphics[width=9cm,bb=0 0 640 480]{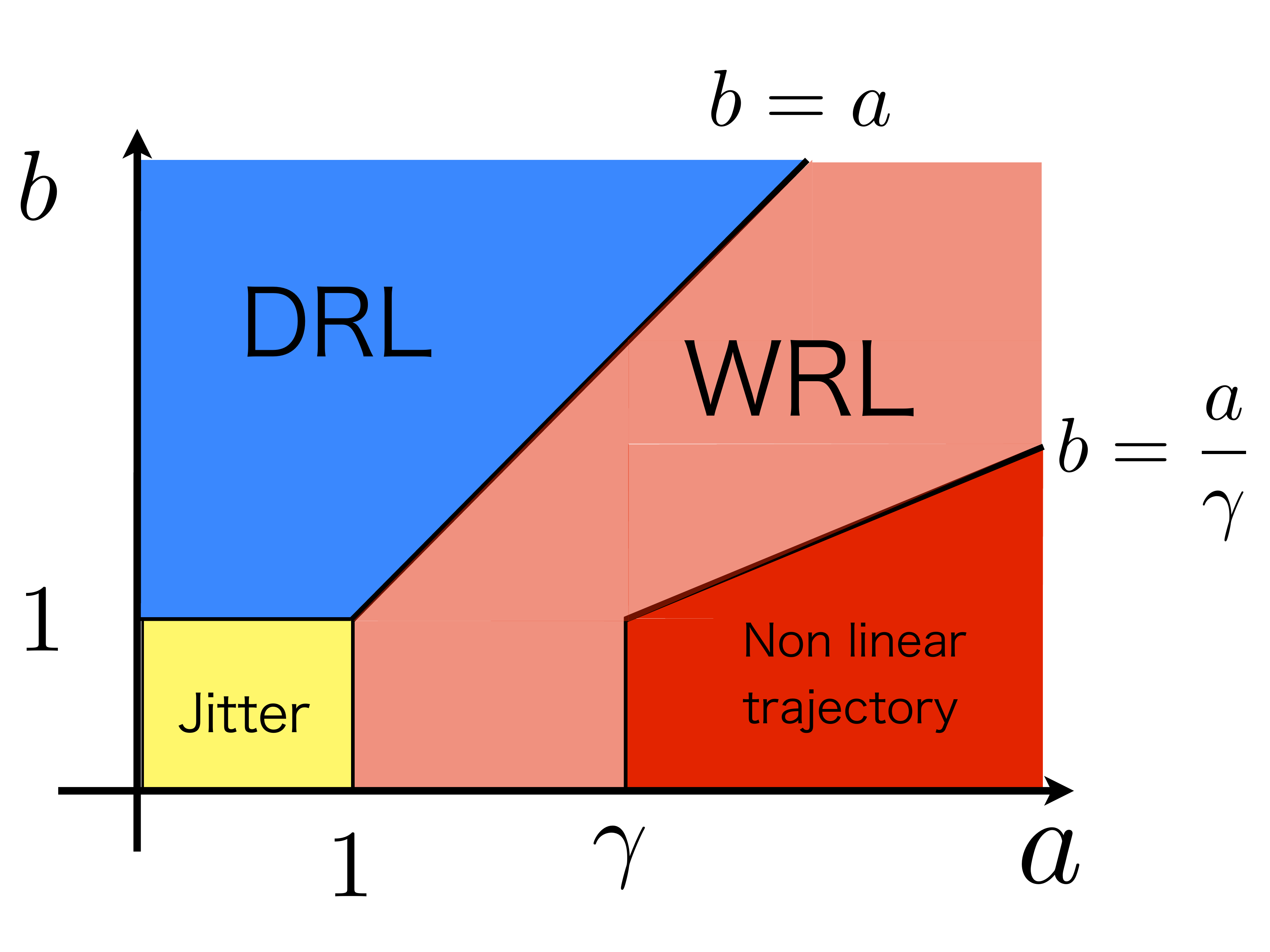}
\caption{Chart of the radiation regimes.
Horizontal axis is $a=\omega_\mathrm{st}/\omega_0=e\sigma/mc^2k_\mathrm{min}$, and
vertical axis is $b=\omega_\mathrm{p}/\omega_0$.
For the jitter regime with $a<1$ and $b<1$, the radiation spectra are determined by the spacial fluctuations,
because $\omega_0$ is the largest of the three.
The typical frequency for this case is $\gamma^2\omega_0$.
For $b>a>1$, i.e., $\omega_\mathrm{p}>\omega_\mathrm{st}>\omega_0$, the radiation spectra are
represented by DRL theory, and typical frequency is $\gamma^2 \omega_\mathrm{p}$.
The line $b=a$ divides the DRL region and WRL region, and the spectral features 
for the WRL regime $a>b>1$ are newly clarified in this paper.
The typical frequency is $\gamma^2\omega_\mathrm{st}$ 
and the spectral index of frequency region lower than the peak is $\sim 1/3$,
in the same way as synchrotron radiation.
For  $a>\gamma$ and $a>\gamma b$, the orbit of a radiating electron depicts non-linear trajectory,
and its signature appears at the low frequency region of the spectrum.
}
\label{a-b_plane}
\end{figure}

We have calculated the radiation spectra from relativistic electrons moving in a Langmuir turbulence 
by using first principle numerical calculation.
We characterize the radiation spectra by two parameters. 
The one is $a=\omega_\mathrm{st}/\omega_0$, where $\omega_\mathrm{st}=e\sigma/mc$ is the strength omega, 
and $\omega_0=2\pi c/\lambda$ is the spatial omega.
The strength omega accounts for the effect of the field strength to the radiation spectra,
and the spatial omega accounts for the effect of spatial fluctuation with a typical scale of $\lambda$.
The other is $b = \omega_\mathrm{p}/\omega_0$, where $\omega_\mathrm{p}$ is the plasma frequency,
which accounts for effects of the time variability of the waves.
We investigate the spectral signatures in the $a-b$ plane (Figure \ref{a-b_plane}).
For $a<1$ and $b<1$, the spectral features are the same as those of jitter radiation or Diffusive Synchrotron Radiation.
For $b>a>1$ and $b>1>a$, the theory of the Diffusive Radiation in Langmuir turbulence is confirmed,
where time variability plays a primary role.
For $a>b>1$, the spectra show novel features which are not predicted by DRL theory.
In this regime, the peak frequency is $\sim\gamma^2 \omega_\mathrm{st}$, 
which is higher than the predicted frequency $\gamma^2\omega_\mathrm{p}$ from the DRL theory.
The spectral index of the frequency region lower than the peak is $\sim 1/3$.
These features are explained by the Wiggler mechanism.
To clarify the radiation features in this regime, we calculate the radiation spectra from an electron moving in an 
oscillating electric field, i.e., for vanishing spatial omega.
We analytically calculate the motion of the electron, and numerically calculate the radiation spectra form this electron.
We show that for $\eta=a/b\gtrsim 1$, the spectrum around the peak frequency consists
of the higher harmonics of the fundamental mode, by considering the radiation in the mean velocity frame.
The electron motion becomes relativistic for $\eta>1$ even in this frame, 
so that strong higher harmonics photons are emitted because of the retarded time effect.
As a result, the spectra in the observer frame consists of the higher harmonics of $\gamma^2\omega_{\rm p}$.
The peak frequency is characterized by $\gamma^2\omega_{\rm st}$,
which is understood by the analogy of the Wiggler radiation.

The feature that the radiation spectra from Langmuir turbulence have a wide range of spectral indices
can be important for high energy astrophysical objects, in particular gamma ray bursts.
The emission mechanism of GRB is not settled for now.
The spectral indices of low frequency side of the Band function are distributed as a Gaussian with the central value of $0$.
Non-negligible number of GRBs have spectral index harder than the theoretical limit for synchrotron radiation $1/3$.
The photospheric emission model can overcome this difficulty, but it also has
a difficulty that intrinsic photosheric emission is too hard with a low energy spectral index of $2$
and it is difficult to make it softer.
On the other hand, the radiation mechanism from Langmuir turbulence in this paper has some advantages.
Not only the spectral index is harder than the synchrotron radiation and 
it can reproduce very hard spectra of observed GRB (Fleishman 2007b),
but also it may explain a wide range of spectral indices.
Because the parameters of $a$ and $b$ are likely to have a value around 1 near the shock front (Silva 2006, Dieckmann 2005),
the radiation spectra change drastically around these parameters.

We are grateful to the referee for his/her constructive and helpful comments
and for informing us of the interesting paper by Getmantsev \& Tokarev (1972).
We also thank K. Toma and S. Kimura for useful discussions. 
This work is supported by the JSPS Research Fellowships for Young Scientists (Y.T., 24593).

\clearpage


\end{document}